\pdfoutput=1

\documentclass{egpubl}
\usepackage{eg2019}

\JournalPaper         % uncomment for final version of Journal Paper
\usepackage[pdftex]{graphicx} \pdfcompresslevel=9

\graphicspath{{images/}}

\usepackage{t1enc,dfadobe}
\usepackage{microtype}

\usepackage{egweblnk}
\usepackage{cite}
\usepackage{mathdots}
\usepackage[inline]{enumitem}

\title[Progressive Refinement Imaging]%
{Progressive Refinement Imaging}

\author[M. Kluge et al.]
{\parbox{\textwidth}{\centering M. Kluge$^{1}$,
        T. Weyrich$^{2}$ 
        and A. Kolb$^{1}$ 
        }
        \\
{\parbox{\textwidth}{\vspace{4.7mm}\centering $^1$Computer Graphics Group, University of Siegen, Germany\\
         \{markus.kluge, andreas.kolb\}{@}uni-siegen.de\\
         $^2$Department of Computer Science, University College London, UK\\
         t.weyrich{@}cs.ucl.ac.uk\vspace{5mm}
       }
}
}
\usepackage{amsmath,amssymb}
\usepackage{color}
\usepackage{xspace}
\usepackage[table,xcdraw]{xcolor}
\usepackage[normalem]{ulem}
\usepackage{bm}
\usepackage{subcaption}
\usepackage[export]{adjustbox}
\usepackage{siunitx}
\usepackage{booktabs}
\usepackage{balance}

\usepackage{hyperref}

\usepackage{threeparttable}

\makeatletter
\DeclareRobustCommand\onedot{\futurelet\@let@token\@onedot}
\def\@onedot{\ifx\@let@token.\else.\null\fi\xspace}
\def\etal{~et~al\onedot}
\def\eg{e.g\onedot} 
\def\ie{i.e\onedot}

\makeatother

\def\clap#1{\hbox to 0pt{\hss #1\hss}}%
\def\initials#1{\protect\clap{\smash{\raisebox{1.4ex}{\tiny{\textsf{\textit{#1}}}}}}}%
\makeatletter
\newcommand{\EDIT}[4][]{\protect\@ifundefined{hidecomments}{%
  \strut{\protect\color{#3}{\hspace{0pt}\protect\initials{#2}\protect\sout{#1}{~#4}}}%
  }{}}
\newcommand{\NOTEboxed}[3]{\protect\@ifundefined{hidecomments}{%
  {\begin{center}\fbox{\parbox{0.97\linewidth}{\protect\EDIT{#1}{#2}{#3}}}\end{center}}
  }{}}
\makeatletter
\newcommand{\DefAuthor}[2] % initials, color
{%
  \expandafter\newcommand\csname #1edit\endcsname[2][]{\protect\EDIT[##1]{#1}{#2}{##2}}
  \expandafter\newcommand\csname #1\endcsname[1]{\protect\csname #1edit\endcsname{[##1]}}
  \expandafter\newcommand\csname #1boxed\endcsname[1]{\NOTEboxed{#1}{#2}{##1}}
}
\makeatother

\newcommand{\bp}[1]{{\ensuremath{\boldsymbol #1}}}

\newcommand{\abs}[1]{ \left| #1 \right| }

\begingroup \catcode0=\active
  \gdef\eightbit#1#2{\catcode`#1=\active
    \begingroup \lccode`\^^@=`#1\relax
    \lowercase{\endgroup \def^^@}{#2}}%
\endgroup

\eightbit{ä}{\"a} \eightbit{Ä}{\"A} \eightbit{ö}{\"o}
\eightbit{Ö}{\"O} \eightbit{ü}{\"u} \eightbit{Ü}{\"U}
\eightbit{ß}{\ss{}}

\def \path{\bp C}

\def \Uo0 {U \backslash \{ 0 \} }
\def \So0 {S \backslash \{ 0 \} }
\renewcommand{\theenumi}  {\arabic{enumi}}

\renewcommand{\theenumii}  {\arabic{enumii}}

\renewcommand{\theenumiii}  {\arabic{enumiii}}

\def \Rom#1 {\uppercase\expandafter{\romannumeral #1}}
\def \rom#1 {\expandafter{\romannumeral #1}}

\def \({\left( }
\def \){\right) }
\def \[{\left[ }
\def \]{\right] }
\newcounter{secrefcnt}
\newcounter{slidecnt}
  {%
    \vfill
  \end{slide}%
  }
  
\makeatother
\def\ignore#1{}

\usepackage{bm}
\definecolor{dkred}         {rgb}{0.8,0,0}
\definecolor{orange}        {cmyk}{0,0.61,0.87,0}
\definecolor{dkviolet}      {cmyk}{0.07,0.90,0,0.34}
\definecolor{dkblue}        {rgb}{0.04,0.18,0.7}
\definecolor{dkgreen}       {rgb}{0.0,0.5,0.0}

\DefAuthor{MK}{orange}
\DefAuthor{TW}{dkviolet}
\DefAuthor{AK}{dkgreen}
\DefAuthor{TODO}{dkred}

\hypersetup{
    pdftitle={Progressive Refinement Imaging},
    pdfsubject={This article has been published in final form at https://doi.org/10.1111/cgf.13808.},
    pdfauthor={Markus Kluge, Tim Weyrich, Andreas Kolb},
    colorlinks=true,
    linkcolor=blue,
    citecolor=blue,
    filecolor=blue,
    urlcolor=blue,
}

\begin{document}

% uncomment for using teaser
%\teaser{

%}

\maketitle
%-------------------------------------------------------------------------

\begin{abstract}
    This paper presents a novel technique for progressive online integration of uncalibrated image sequences with substantial geometric and/or photometric discrepancies into a single, geometrically and photometrically consistent image. Our approach can handle large sets of images, acquired from a nearly planar or infinitely distant scene at different resolutions in object domain and under variable local or global illumination conditions. It allows for efficient user guidance as its progressive nature provides a valid and consistent reconstruction at any moment during the online refinement process.
Our approach avoids global optimization techniques, as commonly used in the field of image refinement, and progressively incorporates new imagery into a dynamically extendable and memory-efficient Laplacian pyramid. Our image registration process includes a coarse homography and a local refinement stage using optical flow. Photometric consistency is achieved by retaining the photometric intensities given in a reference image, while it is being refined. Globally blurred imagery and local geometric inconsistencies due to, \eg motion are detected and removed prior to image fusion.
We demonstrate the quality and robustness of our approach using several image and video sequences, including handheld acquisition with mobile phones and zooming sequences with consumer cameras.
%
%-------------------------------------------------------------------------
%  ACM CCS 1998
%  (see http://www.acm.org/about/class/1998)
% \begin{classification} % according to http:http://www.acm.org/about/class/1998
% \CCScat{Computer Graphics}{I.3.3}{Picture/Image Generation}{Line and curve generation}
% \end{classification}
%-------------------------------------------------------------------------
%  ACM CCS 2012
%The tool at \url{http://dl.acm.org/ccs.cfm} can be used to generate
% CCS codes.
%Example:



\smallskip\medskip 
\end{abstract}
%-------------------------------------------------------------------------

\begin{figure*}[t]
    \centering\scalebox{0.95}{
        \begin{tabular}{@{}l@{}}
          \begin{tabular}{ll@{}}
            \rotatebox{90}{\hspace*{1mm}6 photos} &
            \begin{subfigure}[t]{.793\linewidth}
              \includegraphics[width=1.0\textwidth]{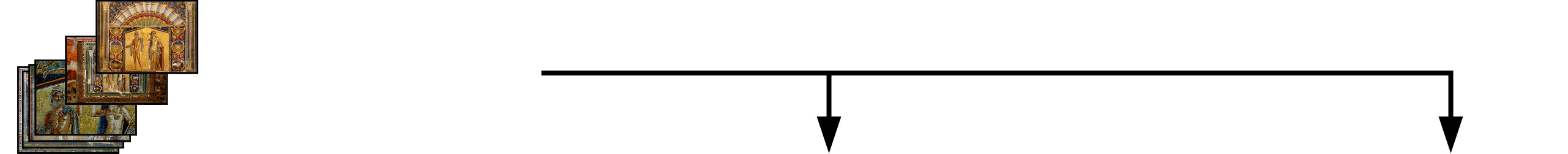}
              \label{fig:results:House_of_Neptune_and_Amphitrite_mosaic.input_frames}
            \end{subfigure}
          \end{tabular} \\  
          \begin{subfigure}[t]{.31\linewidth}
            \includegraphics[width=0.99\textwidth, cfbox=red 1pt 0pt]{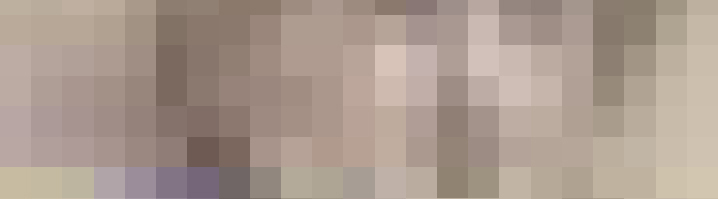}
          \end{subfigure}
          \begin{subfigure}[t]{.31\linewidth}
            \includegraphics[width=0.99\textwidth, cfbox=red 1pt 0pt]{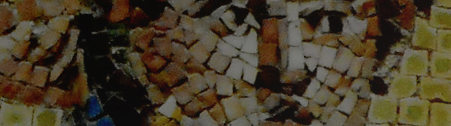}
          \end{subfigure}
          \begin{subfigure}[t]{.31\linewidth}
            \includegraphics[width=0.99\textwidth, cfbox=red 1pt 0pt]{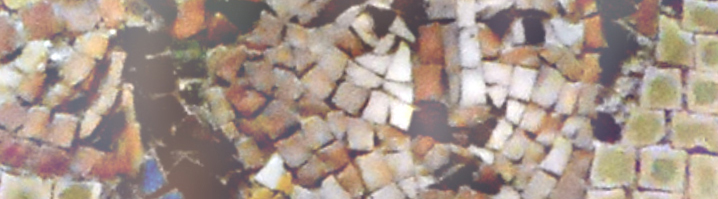}          
          \end{subfigure}\\
          \begin{subfigure}[t]{.31\linewidth}
            \includegraphics[width=1.0\textwidth]{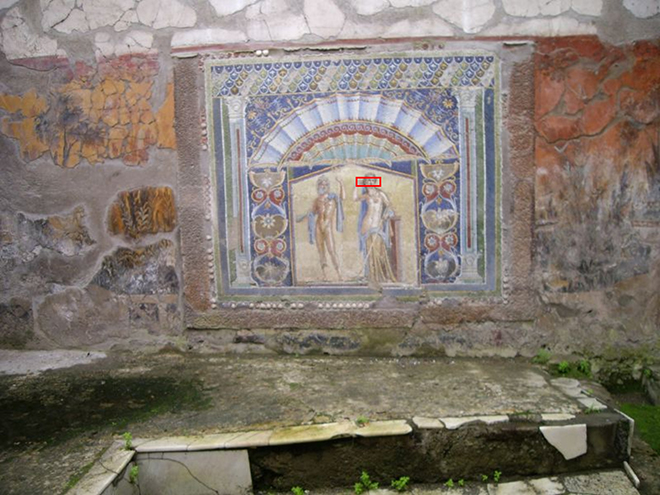}
            \caption{Reference image $\mathcal{I}_0$}
            \label{fig:results:houseOfNeptune.input_1}
          \end{subfigure}
          \begin{subfigure}[t]{.31\linewidth}
            \includegraphics[width=1.0\textwidth]{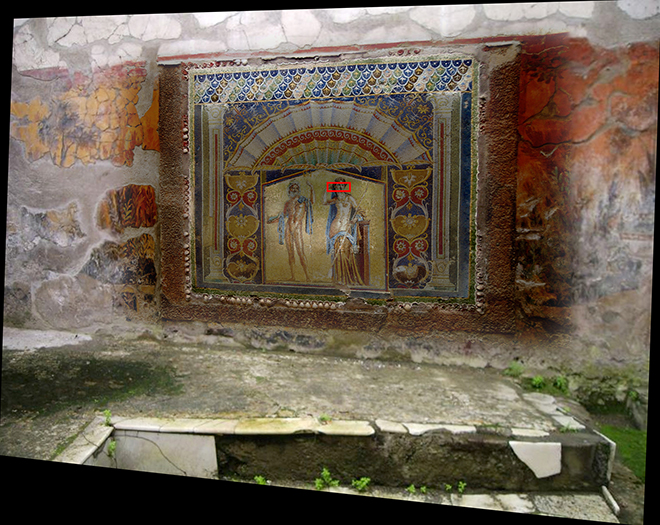}
            \caption{Autopano Giga}
            \label{fig:results:houseOfNeptune.autopano}
          \end{subfigure}
          \begin{subfigure}[t]{.31\linewidth}
            \includegraphics[width=1.0\textwidth]{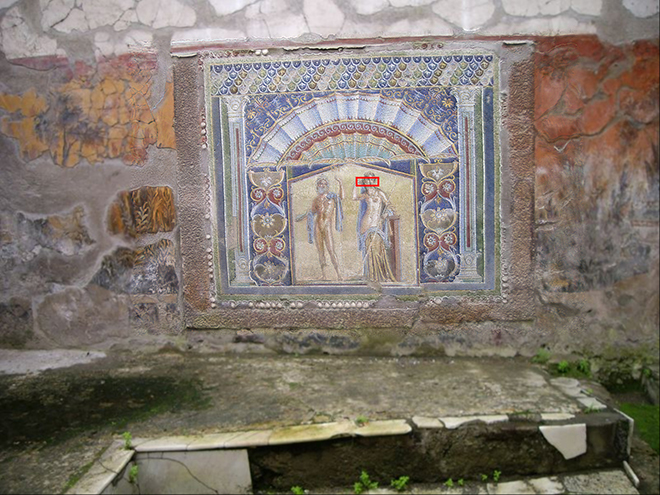}          
            \caption{Ours}
            \label{fig:results:houseOfNeptune.ours}
          \end{subfigure}
        \end{tabular}
    }
    \caption{A sample result of our progressive refinement imaging pipeline applied to the \textit{House of Neptune and Amphitrite mosaic} data set comprising one reference image $\mathcal{I}_0$ that is refined using six additional images captured with six different cameras over the period of 10 years. Compared to prior work, our method successfully generates photometrically and geometrically consistent results in an online and memory-efficient fashion without global optimization. \\}
    \label{fig:results:houseOfNeptune}
\end{figure*}

\section{Introduction}

The visual appearance of real-world objects and scenarios spans
multiple scales, and yet, despite an impressive rise in sensor
resolution, photographic imaging hardware is hardly able to
simultaneously capture visual details across all of these scales.
Several algorithmic approaches have been proposed to overcome the
resolution limits of digital imaging, creating higher resolution
images by fusing information from multiple observations.

%Super-resolution techniques obtain a high-resolution image from multiple low-resolution images~cite{park2003super}, commonly consisting of a registration, interpolation, and final restoration step. They exploit sub-pixel shifts between the individual images and the given (or estimated) point-spread function of the camera in order to solve the related inverse problem by means of global optimization. Super-resolution approaches commonly require a large mutual overlap of the observations, a nearly in-plane motion between images, and strongly rely on sufficient aliasing to be present in the imager.  Not the least due to these many constraints, practical applications are limited to specialized domains where the imaging process meets hard physical limits, such as satellite imaging, microscopy, or computed tomography~cite{nasrollahi2014superresolution}. Moreover, the achievable increase in resolution is limited, typically well below an order of magnitude, and computational costs are high.
Super-resolution techniques obtain a high-resolution image from multiple low-resolution images~\cite{park2003super}, exploiting sub-pixel shifts between the individual images and solving the related inverse problem involving the camera's point-spread function by means of global optimization. Super-resolution techniques are mainly applied to overcome hard physical acquisition limits, such as in satellite imaging, microscopy or computed tomography~\cite{nasrollahi2014superresolution}.

In contrast, computational methods for image recombination and fusion have been developed that address the acquisition of scenes or objects that cannot be captured with a single photograph. Examples are panoramic photography, photo montage~\cite{agarwala2004interactive}, multi-perspective image combination~\cite{yu2008multiperspective} and photo exploration techniques based on partial 3D scene reconstruction from unstructured collections of photographs~\cite{snavely2006photo}. 
Multi-perspective imaging combines images that are acquired under different perspectives using non-standard, potentially non-physical camera models~\cite{yu2008multiperspective} such as computational zoom~\cite{badki2017computational}, which allows modifying image composition parameters, such as the relative magnification of objects or the extent of perspective distortion.

Panoramic photography extends image resolution laterally, by creating a wide-angle mosaic from a set of images with a narrower field of view and small overlapping regions~\cite{szeliski1997creating}. Both alignment and stitching are usually formulated as global optimization problems,  constrained by assuming that all images share the same viewpoint.
The achievable panorama size is generally unlimited and allows for gigapixel imaging~\cite{kopf2007capturing}, while the object-space resolution is determined by the resolution and focal length of the camera used. Alternatively, a low-resolution reference image that completely covers a scene of interest can be enriched with high-resolution details from close-ups~\cite{eisemann2010photo}; our proposed method takes a similar approach.

All methods mentioned above have in common that they process images in batch mode, after capture.  Inspired by progressive acquisition approaches in 3D scene reconstruction~\cite{zollhofer2018state}, we avoid global optimization and super-resolution, and deliberately aim at a progressive framework that allows for continuous addition of observations, resulting in a lightweight and robust image acquisition approach that allows 
\begin{enumerate*}[label=(\arabic*)]
\item unconstrained input imagery, \eg handheld video or mixed-field-of-view images, without requiring calibration, pre-alignment, external tracking, lighting adjustment or other intervention;
\item online user guidance for casual capture and dynamic refinement, even in fleeting situations; and
\item fusing hundreds of images by continuously eliminating redundancy, thus taking the burden of efficiency-conscious view planning from the user.
\end{enumerate*}

Similar to prior work~\cite{eisemann2010photo}, our progressive refinement procedure aims at the addition of high-resolution details to a reference image that covers the region of interest (see Figure~\ref{fig:results:houseOfNeptune}). At the core of our method is an adaptive and expandable Laplacian image-pyramid representation that is used to accumulate further observations into the reference image and which locally increases image resolution and expands the image laterally on demand.
Due to its progressive nature and low costs of decoding, this
representation provides a valid and consistent adaptive-resolution
reconstruction at any moment during the progressive imaging process.
Similar to conventional panoramic imaging, our implementation assumes absence of strong parallax in the input images. However, our approach allows for general camera viewpoints spanning a wide range of resolutions and imagery with strongly varying lens characteristics.

In summary, we propose a simple, still effective approach
to progressively integrate an open set of images into a single
geometrically and photometrically consistent image of a near-planar scenery.
Unique strengths and contributions of our approach are
\begin{itemize}
\item the ability to robustly process uncalibrated, potentially unsharp, geometrically and photometrically inconsistent images at different levels of object resolution and from different viewpoints,
\item the continuous local resolution adjustment to meet the resolution and extent of the incoming images and
\item the scalability into gigapixel range while maintaining near-constant update times upon incoming images.
\end{itemize}

%
%During the reconstruction, incoming source images are locally aligned
%and their high frequencies are added to the model image, if
%appropriate
%
%can handle a large sequence of
%images, \eg, videos, under unconstrained camera motion and varying
%intrinsic (lens) parameters.
%
%This opens up the possibility of online refinement imaging,
%guiding the user in an interactive way to areas where details
%are missing as well as enabling live color reconstructions,
%e.g., for VR applications.
%
%Our approach comprises the following features and contributions:
%\begin{itemize}
%\item progressive... 
%\item handles variable illuminations...
%\item uncalibrated... 
%\item not restricted to a shared viewpoint... 
%\item can handle massive input frames
%\end{itemize}

%-------------------------------------------------------------------------
\section{Related Work}
\subsection{Photo montage}
%%%% Photo montage 

In the mid-19th century, photo montage evolved as a photographic
art form. Rejlander~\cite{rejlander1857twoway}, for example, composed
the allegorical photo `The Two Ways of Life', a photomontage of 32
carefully composed and feathered pictures, and
Robinson~\cite{robinson1869pictorial} discusses principles on how to
arrange form, light and shadow to create the perfect photo
composition in the context of the aesthetics ideal of the
`Picturesque', a concept popularized in the mid-18th century.
Today, applications of photo montage have gone well beyond the artistic
medium, and digital workflows employ modern-day equivalents that build
upon works such as digital image mosaicing~\cite{milgram1975computer}
and photomontage~\cite{agarwala2004interactive}.

In the digital domain, the main technical challenge is to recombine images without leaving visible traces at the seams where images are composited. Previous works explored strategies for visually least disruptive placement of seams~\cite{milgram1975computer,efros2001image,kwatra2003graphcut,agarwala2004interactive,li2004lazy} and blending operations to obscure image differences across a seam, such as linear feathering~\cite{milgram1975computer}, Poisson blending~\cite{he2017gigapixel,szeliski2011fast,pulli2010mobile,agarwala2004interactive} and the multi-resolution spline approach~\cite{burt1983multiresolution} that gave rise to the Laplacian image pyramid~\cite{burt1984pyramid,Ogden1985pyramid}.
Laplacian image pyramids allow for computationally efficient multi-scale image representation in a localized, frequency-oriented way~\cite{adelson1984pyramid,paris2011local}.
Burt and Adelson~\cite{burt1983multiresolution} were the first to fuse images generating smooth transitions by using Laplacian pyramids and spatial blending. Burt and Kolczynski~\cite{burt1993enhanced} extend this idea by addressing the objective of combining several, pre-aligned source images into a single composite image retaining specific image regions while discarding other image portions.
%
%A hugely influential paper, giving rise even to recent works, such as
%exposure fusion~\cite{mertens2007exposure}, it also inspired our own
%work.

%% Kwatra\etal\cite{kwatra2003graphcut} introduced graph-cuts to
%% image fusion in the context of texture synthesis. This
%% approach can also compose two source images into a single
%% destination image by finding optimal cuts in the source
%% images.  Agarwala\etal\cite{agarwala2004interactive} extend
%% this technique by allowing to combine several, sequentially
%% taken images into a single seamless photo. The images are
%% taken from (more or less) the same scene and they need to have
%% sufficiently consistent object resolution and illumination
%% conditions. By adding strokes on the input images, the user
%% can specify regions of interest that should be fused into a
%% single image. Applying a graph-cut optimization, good seams
%% are identified that allow a seamless result that is
%% generated using Poisson-based gradient-domain fusion.

%\TODO{newer papers on photo montage?}

\subsection{(Very large) Panoramic images}
%%%% Panoramic images 

Panoramic photography is strongly related to seamless photomontage, as it attempts to combine several images into a consistent, artefact-free image. Geometric registration is facilitated via feature matching, either based on simple landmarks~\cite{milgram1975computer} or on more complex features like SIFT~\cite{brown2007automatic}. For image composition, blending strategies including Poisson, Laplacian and multi-band blending are used~\cite{szeliski1997creating,brown2007automatic,pulli2010mobile,szeliski2011fast,he2017gigapixel}.
%
%\TODO{Give an overview on image stitching approaches \ldots}

Kopf\etal\cite{kopf2007capturing} introduced a system to acquire gigapixel images, \ie wide angle images of extremely high resolution.
Their source imagery consists of robotically captured, geometrically uncalibrated high dynamic range (HDR) image stacks that are automatically undistorted using feature matching. Overall geometric consistency is achieved via global bundle adjustment. Photometric consistency results from an exposure adjustment utilizing the linear intensity domain of the HDR imagery and a photometric alignment and composition technique~\cite{eden2006seamless}. The final composition is achieved using a graph-cut.
Kazhdan and Hoppe~\cite{kazhdan2008streaming} proposed new methods for editing gigapixel images. Their out-of-core multi-grid approaches allows for gradient-domain image-editing operations involving the solution of Poisson equations that exceed the main memory capacity in the case of gigapixel images.
Follow-up work on gradient-domain editing of gigapixel images extends the gigapixel approach towards wide-angle, high-resolution looping panoramic videos synthesis~\cite{he2017gigapixel}.

In our work, these challenges do not occur, as our blending operation
takes place directly on the hierarchical Laplacian representation.

\subsection{Photo collections}

Several works have extended the idea of panoramic photography to more
general image sources.
Snavely\etal's Photo Tourism system~\cite{snavely2006photo} processes unstructured photo collections of popular internet sites, taken with various different cameras, at different times of the day, different seasons or from various unknown positions. Instead of generating a single output image, their system merely recovers the camera poses and a sparse point cloud, and offers a 3D interface to browse through these photographs within their 3D context.
Similarly, Ballan\etal\cite{ballan2010unstructured} source both still images as well as handheld videos to create a browsable 3D representation that embeds original camera views in a rough 3D spatially and temporally synchronized reconstruction of the event.
While these works circumvent the challenge of creating a seamless reconstruction, the use of unstructured collections of photographs, similar to our approach, requires robust alignment of uncalibrated photographs.
Further work in this direction demonstrates the exploration
of video collections within the panoramic context of the same place~\cite{tompkin2013video} and the embedding of
video clips within gigapixel scale imagery~\cite{pirk2012video}.

Eisemann\etal's Photo Zoom~\cite{eisemann2010photo} pursues a similar goal to ours, automatically constructing a high-resolution image from an unordered set of zoomed-in photos, but requires global, post-capture processing. Furthermore, they
\begin{enumerate*}[label=(\arabic*)]
\item tackle colour inconsistencies using a recursive gradient domain fusion approach that cannot handle strong local variations such as reflections,
\item only apply homographies to register images and mask out regions with inconsistent content,
\item expect all input images to be focused and
\item only fuse a comparable small number of images.
\end{enumerate*}
On the flip side, their system synthesizes detail in undersampled regions.

\subsection{Progressive reconstruction}

In a sense, our solution falls into the class of simultaneous localization and mapping (SLAM) algorithms that gradually build up a world model while reconstructing sensor location and orientation (in our case a camera pose) by relating any observations to the model built up so far~\cite{thrun2005probabilistic,newcombe2010live}. Many of these methods share a feature detection and matching stage, similar to the one employed by our method.
Apart from that, a multitude of works combines sensors that range from laser range scanners, through 2D cameras, to handheld depth cameras and merge their observations into various types of environment models (sparse features~\cite{pumarola2017plslam}, collections of range maps~\cite{newcombe2010live}, volumetric grids~\cite{izadi11kinectfusion,niessner2013realtime}, oriented points~\cite{keller2013pointbased}, to name a few). To our knowledge, however, none of these works involves direct updates of an unbounded multi-scale world representation.

\renewcommand{\arraystretch}{1.1}

\begin{table}[b!]
  \small
  \caption{\label{tab:symbols}List of conventions.}
  \begin{tabular*}{\linewidth}{@{}l@{~~~}p{0.73\linewidth}@{}}
    \toprule
    $\mathcal{I}_j$ & $j$th input image, whereas $\mathcal{I}_0$ is the reference image and \newline \hspace*{0.65em} $\mathcal{I}_j,\,j>0$ an observation\\
    $\mathcal{M}$ & Model (refined reference image)\\
    $\mathcal{I}_j^{l}$, $\mathcal{M}^{l}$ & $\mathcal{I}_j$ and $\mathcal{M}$ decomposed in Laplacian pyramid levels \newline \hspace*{0.65em} $l\in[l_{\text{min}}^{\mathcal{I}_j},\,l_{\text{max}}^{\mathcal{I}_j}]$ and $l\in[l_{\text{min}}^\mathcal{M},\,l_{\text{max}}^\mathcal{M}]$, respectively\\
    $l_i^{\mathcal{I}_j}$, $l_i^{\mathcal{M}}$ & Level with a specific scale factor with respect to $\mathcal{I}_0$, \newline \hspace*{0.65em} where $i$ is the level's index in the pyramid\\
    $T^{\mathcal{I}_j}_{(p,\,q),\,l}$, $T^{\mathcal{M}}_{(p,\,q),\,l}$ & $\mathcal{I}_j^{l}$ and $\mathcal{M}^{l}$ split into tiles with 2D array position $(p,\,q)$\\
    $c_{\mathcal{I}_j}^{l}$, $c_{\mathcal{M}}^{l}$ & Confidence map of $\mathcal{I}_j^{l}$ and $\mathcal{M}^{l}$\\ 
    $\mathcal{F}_{\mathcal{I}_j}$, $ \mathcal{F}_{\mathcal{M}}$ & Local feature set in $\mathcal{I}_j$ and $\mathcal{M}$\\
    $\mathcal{H}_j$ & Homography warping  $\mathcal{I}_j$ to $\mathcal{M}$\\
    $\mathcal{L}_j$ & Level map of $\mathcal{I}_j$ storing real-valued level numbers per \newline \hspace*{0.65em} pixel with respect to the model pyramid\\
    \bottomrule
  \end{tabular*}\\[-0.5\baselineskip]
\end{table}

%% Let's not have a "conclusion" to the related work section, but
%% rather establish context at the end of each subsection:
%%
%% \TWboxed{%
%% %Conclusion (hopefully):
%% %
%% All the panoramic image mosaicing
%% techniques acquired with rather constant object resolution,
%% \ie with fixed camera intrinsics\TW{not correct} and constant camera-to-object
%% distance\TW{does not hold for Tomkin et al., for instance}, use a rather heavy-weight acquisition setup that
%% provides high quality source imagery and/or apply global
%% optimization to the complete set of source images.}

%-------------------------------------------------------------------------
\section{Overview}\label{sec:pipeline}

\begin{figure}[tb]
    \centering
    \scalebox{0.7}{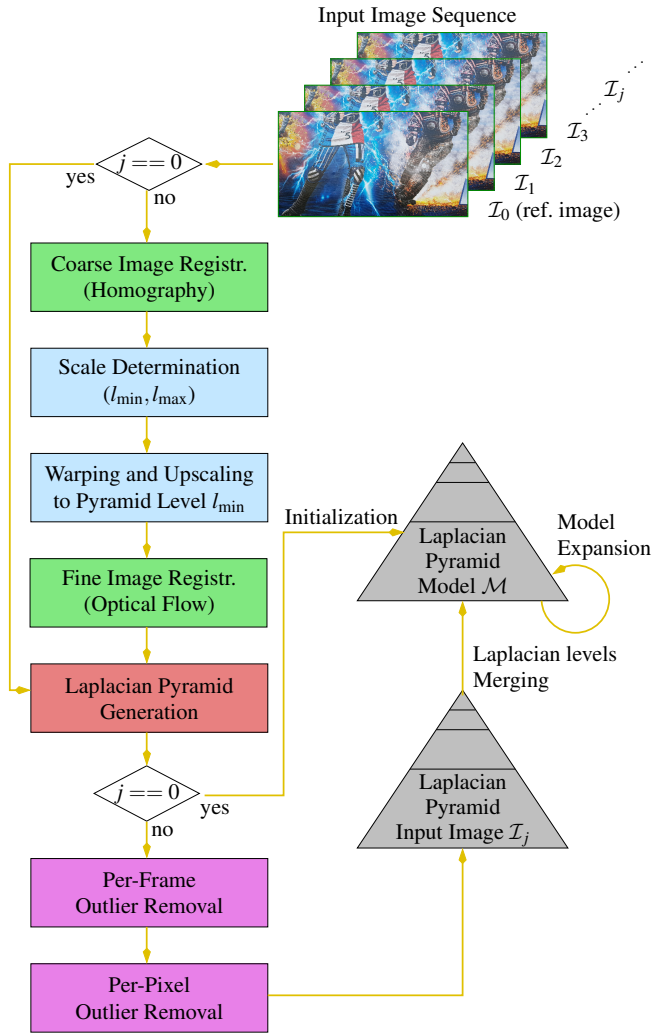}
    \caption{Our progressive refinement imaging pipeline.}
    \label{fig:pipeline}
\end{figure}

Our proposed refinement pipeline comprises several processing stages as depicted in Figure~\ref{fig:pipeline}. We expect the first input image $\mathcal{I}_0$ fed into our pipeline to be a \emph{reference image}, covering the region of interest for all following input images $\mathcal{I}_j,\,j>0$. Within this region initialized by $\mathcal{I}_0$, our system results in a geometric and photometric consistently refined image representation. In the following, we call this representation \emph{model} $\mathcal{M}$. Outside of the region defined by the reference image, we still achieve geometric but no photometric consistency.  See Table~\ref{tab:symbols} for a complete list of conventions used.

The main stages of our pipeline can be summarized as follows:

\subsection{Image registration} 
While the reference image's viewing
direction defines the default view for the refinement process, further
observations $\mathcal{I}_j,\,j>0$ can be acquired from different positions and
viewing directions. To match the model's pixel grid, we
perform an image registration first. This is done by aligning the
observation globally using a homography estimated with the help of
local features.  Afterwards, we locally fine-correct the registration
based on an estimated flow field
(see Section~\ref{sec:progressive-refinement:registration}).

\subsection{Laplacian pyramid generation}
In this pipeline stage, the
registered observation $\mathcal{I}_j$ is decomposed into Laplacian pyramid levels $\mathcal{I}_j^{l_i}\in[{\mathcal{I}_j^{l_{\text{min}}},\ldots, \mathcal{I}_j^{l_{\text{max}}}}]$ that will be (potentially) merged with their corresponding Laplacian model levels $\mathcal{M}^{l}$. These levels are generated by differences of low-pass filtered and downscaled versions of $\mathcal{I}_j$ using the Gaussian-like kernel $[\text{0.0625 0.25 0.375 0.25 0.0625}]$ in 1D~\cite{burt1984pyramid}. Thus, each level contains the frequencies of a specific band. Depending on the viewing direction and position, the Laplacian observation level $\mathcal{I}_j^l$ may contribute to the corresponding model level $\mathcal{M}^l$ by adding new information in several ways. They can provide
\begin{enumerate*}[label=(\arabic*)]
\item high frequencies not present in the model so far,
\item lower frequencies already present, but with less
  precision and/or
\item new spatial coverage not observed so far
\end{enumerate*} (see
Section~\ref{sec:progressive-refinement:extract}).

\subsection{Outlier removal} 
As an incoming observation $\mathcal{I}_j$ may have different deficiencies, we conduct a two-level outlier removal. Firstly, we apply a global reliability check to make sure that $\mathcal{I}_j$ provides valuable frequency information that is consistent with the so far accumulated model $\mathcal{M}$, or if it is out of focus, \eg due to an incorrect autofocus or motion artefacts. On the second outlier removal stage, we compute a pixelwise error on the Laplacian level to recognize local registration errors due to, \eg inaccuracies in the optical flow estimation (Section~\ref{sec:progressive-refinement:clean-observation}).

\subsection{Model expansion} 
We do not restrict the accumulation of observations into the model in terms of scale, resolution or coverage in object domain. Our model representation is an adaptive Laplacian pyramid that can be expanded in both resolution and lateral dimensions to incorporate novel information in either of these directions. Our Laplacian pyramid model $\mathcal{M}$ comprises an adaptive tile-based representation in which tiles are allocated on-demand (see Sections~\ref{sec:representation} and \ref{sec:progressive-refinement:extract}).

\subsection{Merging Laplacian levels} 
At the core of our technique lies the merging of specific Laplacian levels $l_{\text{min}},\ldots, l_{\text{max}}$ of the current observation $\mathcal{I}_j$ and the model $\mathcal{M}$ that depends on specific resolution and/or lateral information provided by $\mathcal{I}_j$. Merging Laplacian levels is based on per-pixel confidence values $c^l_{\mathcal{I}_j}(x,y)$ for the Laplacian levels of $\mathcal{I}_j$ and the corresponding model values $c^l_{\mathcal{M}}(x,y)$. By comparing these confidence values, we are able to decide which pixels are capable of refining our model and how the observation and the model pixel values of the Laplacian levels are combined (see Section~\ref{sec:progressive-refinement:merge}). Note that we never merge the top Gaussian levels of the model and the observation pyramid, but only Laplacian levels, thus retaining global photometric consistency.

Optionally, we render a visualization to steer the user towards image
areas that need further refinement according to his or her needs and
interests (see Section~\ref{sec:progressive-refinement:visualize}).

%-------------------------------------------------------------------------
\section{Adaptive Model Representation}
\label{sec:representation}

\begin{figure}[t!]
  \centering
  \scalebox{0.7}{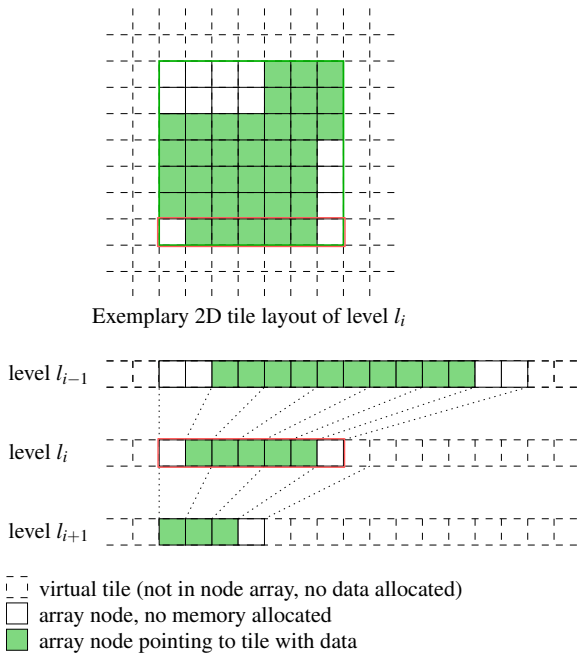}
  \caption{Adaptive Laplacian pyramid. Top: On each pyramid level, a virtually infinite tile array is set up. The nodes in the array form the bounding box (green box) of potential tiles (white squares) and, if required, allocated tiles (green squares). Bottom: Corresponding tiles related to the tile row marked in orange on different pyramid levels (as 1D layout), where two neighbouring tiles are downsampled to a single tile.}
  \label{fig:adaptive-tiled-pyramid}
\end{figure}

Our preliminary goal is to progressively refine a given model image
$\mathcal{M}$ by new input images (observations) $\mathcal{I}_j$ that can be
taken at different scales or resolutions in the object domain and that
cover potentially different regions. Thus, instead of using a flat
representation, an adaptive Laplacian pyramid is an appropriate
representation for our model $\mathcal{M}$. Our adaptive Laplacian
pyramid efficiently stores the model by means of localized detail
information at different resolutions stored in Laplacian levels.
Provided that two images (the observation and the
model image in our case) are properly registered, Laplacian pyramids
offer the advantage of directly comparing and manipulating detail
information on corresponding resolution levels without the computational
burden of an explicit frequency analysis; see
Burt\etal\cite{burt1983multiresolution} for further technical details.

\subsection{Initialization}
Generating the standard Laplacian pyramid for
the initial reference image $\mathcal{I}_0$ defines the initial model $\mathcal{M}$, and thus, serves as a reference view onto the scene. Pyramid level $l^{\mathcal{M}}_i$ describes a model level with a specific scale factor with respect to $\mathcal{I}_0$, where $i$ is the level's index in the pyramid. Index $i=0$ refers to the full resolution of $\mathcal{I}_0$, whereas levels $l^{\mathcal{M}}_i$ with $i>0$ and $i<0$ contain coarser and finer image resolutions, respectively (see Figure~\ref{fig:adaptive-tiled-pyramid}). From level $l^{\mathcal{M}}_i$ to $l^{\mathcal{M}}_{i+1}$, the resolution decreases by one octave, \ie if level $l^{\mathcal{M}}_0$ is defined as sampling distance $1$, level $l^{\mathcal{M}}_i$ has sampling distance $2^i$. All further incoming observations that are potentially acquired from different positions under different view directions are warped appropriately to match this reference view.

\subsection{Adaptivity}
As our model has to be dynamically expanded in
order to represent so far unobserved content, \ie higher or lower Laplacian levels or new lateral regions, we use a tile-based representation of our Laplacian pyramid. As storing a complete Laplacian pyramid would be extremely memory inefficient, we set up a simple regular grid per pyramid level and a 2D node array covering the bounding box of the tiles. While tiles with data are stored in an unordered list, the 2D node array stores the actual layout of the tiles forming a pyramid level of model $\mathcal{M}$. A node points either to the allocated data of its tile or stores $-1$ if no memory is allocated so far. This 2D node array can be extended in lateral direction and new levels can easily be added to represent new resolution levels (see Figure~\ref{fig:adaptive-tiled-pyramid}). New tiles get allocated and assigned to the virtual nodes on demand. We use tiles of size $512\times512$ pixels.

\subsection{Confidence maps}
We log the confidence of the accumulated
model pixels $\mathcal{M}^l(x,y)$ by storing pixelwise confidence
values $c^l_{\mathcal{M}}(x,y)$ for each Laplacian model level $l$. Together with
the confidence values $c_{\mathcal{I}_j}^l(x,y)$ computed for the current observation $\mathcal{I}_j$,
the model's confidence values determine the merging
result (see
Section~\ref{sec:progressive-refinement:merge}).

%-------------------------------------------------------------------------
\section{Progressive Refinement}
\label{sec:progressive-refinement}

Our progressive refinement pipeline uses the Laplacian pyramid of the first input image $\mathcal{I}_0$ of our image sequence as initialization of the model $\mathcal{M}$ (see Section~\ref{sec:representation}). This first input image defines the reference view and the region of interest of the observed scene. Following observations $\mathcal{I}_j$ are integrated if they provide further information in terms of finer details or new lateral image regions. To simplify notation, we omit frame index $j$ in the following, \ie the current observation $\mathcal{I}_{j},\,j>0$ is denoted by $\mathcal{I}$.

%-------------------------------------------------------------------------
\subsection{Image registration}
\label{sec:progressive-refinement:registration}

\begin{figure}[t!]
  \centering
  \def\svgwidth{1.0\linewidth}
  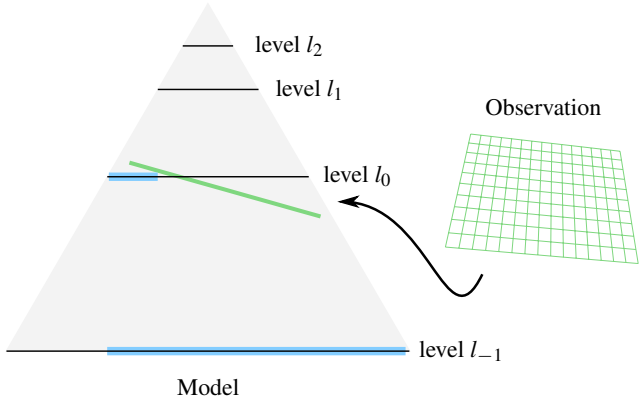
  \caption{An observation is positioned within the adaptive Laplacian
    model pyramid. The observation pixels are warped to the next lower
    corresponding level to match its pixel grid (covering the blue
    marked areas). Thus, the observation
    contributes high frequencies to the Laplacian model level
    $l^{\mathcal{M}}_{0}$ and the new level $l^{\mathcal{M}}_{-1}$ that our model pyramid will adapt
    to.}
  \label{fig:positioning-within-pyramid}
\end{figure}

As we expect the current observation $\mathcal{I}$ to be captured with a different focal length and/or from a different camera pose than the reference view of model $\mathcal{M}$, we first estimate the homography between $\mathcal{I}$ and $\mathcal{M}$. Therefore, we detect a set of local features $\mathcal{F}_\mathcal{I}$ in $\mathcal{I}$ and use the so far accumulated model features $\mathcal{F}_{\mathcal{M}}$, detected in previous observations. Each set $\mathcal{F} = \{ ({x_k,y_k}, f_k ) \; | \; k = 1,\ldots, n\}$ of $n$ detected features is defined by its position ${x_k,y_k}$ and its descriptor $f_k$. In our pipeline, we use speeded-up robust features (SURF)~\cite{bay2006surf} as it provides a fast and robust detection. The homography $ \mathcal{H}$ is estimated by applying a RANSAC matching~\cite{fischler1981random} to the feature sets $ \mathcal{F}_\mathcal{I} $ and $ \mathcal{F}_{\mathcal{M}}$.  As we assume some spatial coherence between consecutive input images, which is especially true in case of video sequences, we use the homography of the previous frame as initialization. To accumulate features for later usage without having to reconstruct the model pyramid, we replace all features $\mathcal{F}_{\mathcal{M}}$ positioned within the observed area by new features $\mathcal{F}_\mathcal{I}$, if the observation passes the full image outlier check in Section~\ref{sec:progressive-refinement:clean-observation}. Since all positions $(x_k,y_k)$ of $\mathcal{F}_{\mathcal{M}}$ are related to the finest model level $l^{\mathcal{M}}_{\text{min}}$, we transform the positions of $\mathcal{F}_\mathcal{I}$ accordingly. This re-positioning is also performed on $\mathcal{F}_{\mathcal{M}}$ after the model gets extended to finer levels.

Using the homography $\mathcal{H}$, we now position the observation $\mathcal{I}$ with respect to lateral and (real-valued) level position in the model pyramid (see Figure~\ref{fig:positioning-within-pyramid}). This yields the minimal and maximal levels $l_{\text{min}}, l_{\text{max}}$ in the model pyramid that bound the scale of $\mathcal{I}$.  As we want to avoid information loss due to downsampling, we warp the observation to the corresponding pyramid level $l_{\text{min}}$ (\eg level $l^{\mathcal{M}}_{-1}$ in Figure~\ref{fig:positioning-within-pyramid}). To maintain the original level positioning, we compute a corresponding level map $\mathcal{L}$ by storing the real-valued level number with respect to the model per pixel (see also Section~\ref{sec:progressive-refinement:merge}).

As we take uncalibrated observations as input, we expect mismatches especially in border and corner regions applying the homography only. To reduce this mismatch to a minimum, we fine-correct the registration locally. To achieve this, we need to compute the displacement for each pixel of $\mathcal{I}$ so that the photometric consistency between $\mathcal{I}$ and $\mathcal{M}$ of the observed area is as high as possible.  A dense optical flow~\cite{horn1981determining, lucas1981iterative} estimates the pixelwise motion between two frames, resulting in a 2D flow field that contains the required displacement vectors.  Therefore, we perform a backward optical flow between $\mathcal{M}$ and $\mathcal{I}$ of the observed area on level $\text{max}(l_{\text{min}}, l^{M}_{\text{min}})$, where $l^{M}_{\text{min}}$ is the lowest level before the model expansion. After potentially resizing the flow field to level $l_{\text{min}}$, we resample $\mathcal{I}$ accordingly. In our implementation, we use an OpenCV function with GPU acceleration that implements an optical flow variant presented by Farneb\"ack\etal~\cite{farneback2003two}.

%-------------------------------------------------------------------------
\subsection{Generation of the Laplacian pyramid}
\label{sec:progressive-refinement:extract}

Considering $\mathcal{M}^{l}$ and $\mathcal{I}^{l}$, the Laplacian pyramids of the model and the warped observation,
their finest levels are defined by $l_{\text{min}}^{\mathcal{M}}$ and $l_{\text{min}}^\mathcal{I}$,
whereas $l_{\text{max}}^{\mathcal{M}}$ and $l_{\text{max}}^\mathcal{I}$ are the coarsest levels. Since we generate
a new pyramid for each observation, $l_{\text{min}}^{\mathcal{I}}=l_{\text{0}}^{\mathcal{I}}$ always holds,
and the corresponding levels in the adaptive model pyramid are defined by the same scale in object domain
(\eg in case of Figure~\ref{fig:positioning-within-pyramid}, $l_{-1}^{\mathcal{M}}$ and
$l_{0}^\mathcal{I}$ are corresponding levels). Furthermore, we have allocated model and observation
tiles $T^{\mathcal{M}}_{(p,\,q),\,l}$ and $T^\mathcal{I}_{(p,\,q),\,l}$, where $(p,\,q)$
is the tile's position in the 2D tile array and $l$ the pyramid level with
$l\in[l_{\text{min}}^{\mathcal{M}},\,l_{\text{max}}^{\mathcal{M}}]$ for model
tiles and $l\in[l_{\text{min}}^\mathcal{I},\,l_{\text{max}}^\mathcal{I}]$ for observation
tiles. When capturing the scene from different positions, an observation can
contribute content for merging into the model considering three
cases:

\paragraph*{Contributing finer image information.} 
The new observation shows the scene captured from a closer distance, \eg after moving the camera towards the scene or zooming in. In this case, some observation tiles $T^\mathcal{I}_{(p,\,q),\,l}$ are not yet in the model pyramid, but corresponding tiles on coarser levels are. Thus, we extract the required tiles of the Laplacian level from the observation and add them to the model pyramid. As observation tiles also contribute to already existing model tiles, a merging of the model and the observation is applied in this case (see Section~\ref{sec:progressive-refinement:merge}).

\paragraph*{Contributing new scene areas at existing pyramid levels.} 
The observation may provide new areas outside the current image boundaries, which allows more of the scene to be included in the reconstruction. In this case, we use all pyramid levels up to $l_{\text{max}}^{\mathcal{M}}$ for incorporation into our model representation. Tiles that are not present in the model will be added, existing tiles will be merged (see Section~\ref{sec:progressive-refinement:merge}). Note that in this situation, photometric inconsistencies may occur on the top Gaussian level of the model pyramid outside of the region defined by the reference image $\mathcal{I}_0$.

\paragraph*{Contributing coarser image information.} 
Similar to the prior case, moving the camera farther away or zooming out results in newly observed regions, but also in coarser Laplacian levels not yet present in the model, \ie $l^\mathcal{I}_{\text{max}}>l^{\mathcal{M}}_{\text{max}}$. Thus, we additionally have to add higher pyramid levels into our model. In this case, we expand the model's Laplacian pyramid to the same level as the one of the observation, \ie to $l^\mathcal{I}_{\text{max}}$. Again, as in the prior case, photometric inconsistencies may occur on the top Gaussian level of the model pyramid.

%-------------------------------------------------------------------------
\subsection{Outlier removal}
\label{sec:progressive-refinement:clean-observation}

Before merging the Laplacian levels of the current observation $\mathcal{I}$ into our model pyramid, we apply an outlier removal in a full image and in a per-pixel stage. Here, outlier refers to image details of the observation $\mathcal{I}$ that are inconsistent to the so far accumulated model $\mathcal{M}$, and thus, should not be merged into our model. The main reasons for photometric inconsistencies are out-of-focus or motion blurred images that should be rejected completely, and local inconsistencies due to inaccurate flow estimations or dynamic scene parts (see Section~\ref{sec:progressive-refinement:registration}).

\subsubsection{Full image outlier} 
We check for global consistency by comparing the finest Laplacian levels of the warped observation $\mathcal{I}$ and the model $\mathcal{M}$. Here, we apply a simple rule assuming that the novel observation contains at least as many fine details as the current model. Therefore, we compute the standard deviation of $\mathcal{I}^{l}$ and $\mathcal{M}^{l}$ on Laplacian level $l_{\text{min}}^\mathcal{I}$. If the standard deviation of the observed Laplacian level is smaller than the model values, we conclude that the observation does not provide additional image details and we drop $\mathcal{I}$.

\subsubsection{Per-pixel outlier}
If the observation $\mathcal{I}$ passed the full image outlier check, we compute a per-pixel matching error that accounts for imperfect local warps due to flow estimation insufficiencies or to dynamic scene parts. As local error metric, we use the relative absolute error $E(x,y)$ on Laplacian levels $l\in[l_{\text{min}},\,l_{\text{max}}[$, with $l_{\text{min}}:=\text{max}(l^\mathcal{M}_{\text{min}},\,l^\mathcal{I}_{\text{min}})$ and $l_{\text{max}}:=l^\mathcal{I}_{\text{max}}$. Note that we exclude the top Gaussian level $l_{\text{max}}$ for comparison due to its susceptibility to false positives if local photometric inconsistencies exist between $\mathcal{I}$ and $\mathcal{M}$. Moreover, in order to reduce the effect of considering new incoming details as outliers, we do not include high-frequency levels that are only present in $\mathcal{I}$, as $l_{\text{min}}$ is the finest level that exists in both pyramids. The per-pixel error is computed as
$$ 
E(x,y) = \sum_{l\in[l_{\text{min}},\,l_{\text{max}}[} \frac{\abs{\mathcal{M}^{l}(x,y) -    \mathcal{I}^{l}(x,y)}}{\text{min}(\abs{\mathcal{M}^{l}(x,y)},\,\abs{\mathcal{I}^{l}(x,y)})}.
$$
In all our experiments, we discard observation pixels with $E(x,y)>10$ in the case of low geometric distortions and with $E(x,y)>1$ in the case of strong geometric distortions, \ie for data sets \textit{Moving cars} in Figure~\ref{fig:results:Moving_Cars} and \textit{Streetart fisheye} in Figure~\ref{fig:results:Streetart_Fisheye}. The idea behind this decision is that the model contains consistent detail information across the Laplacian levels. The error will become large, if the observation adds specifically high values in areas, where the model contains very small values only, or vice versa. This is a clear indication that the observation is locally inconsistent. For reasons of noise removal and filling in gaps, the resulting mask is post-processed by a morphological opening followed by a closing. For these operations we use a disk-shaped structuring element with radius $r = 3$ pixels and $r = 4$ pixels, respectively. If the observation contributes new image regions, and thus, the model does not contain any data, we add the observation content anyway.

%-------------------------------------------------------------------------
\subsection{Merging of model and observation Laplacian levels}
\label{sec:progressive-refinement:merge} 

In the following, we consider individual pixels $\mathcal{M}^l(x,y)$ in the Laplacian model pyramid at level $l$ that already contain data and for which we have observation pixels $\mathcal{I}^{l}(x,y)$ that need to be merged, \ie the pixels have passed the outlier test (see Section~\ref{sec:progressive-refinement:clean-observation}). Furthermore, we have the level map $\mathcal{L}$ that contains the real-valued level numbers of the pixels of $\mathcal{I}$ with respect to the model pyramid levels (see Section~\ref{sec:progressive-refinement:registration}). 

%\begin{figure}[!]
%    \centering
%    \def\svgwidth{1.0\linewidth}
%    \input{images/confidence-assignment.pdf_tex}
%    \caption{Model confidence values for the replacement strategy for three sample pixels. The orange, pink, and green observation pixel have been captured with level map values, \ie, real-valued level numbers, $\mathcal{L}(x,y)=0.6, 1.0,$ and $1.1$, respectively. Newly incoming pixels will replace the model pixel, if their confidence value (exact fractional level) is smaller and, thus, provides more reliable image detail.}
%    \label{fig:confidence-assignment}
%\end{figure}

Inspired by online 3D scene reconstruction~\cite{zollhofer2018state}, we additionally compute confidence values $c^l_\mathcal{I}(x,y)$ for the Laplacian observation levels $l$ of $\mathcal{I}^{l}$ that refer to the reliability of the pixels $\mathcal{I}^{l}(x,y)$. The model confidence values are stored in $c^l_{\mathcal{M}}$ for $\mathcal{M}^{l}$. 
In the case of image fusion, we relate the confidence to the contrast in a focused image, which can be measured using the modulation transfer function (MTF) of a camera; see, for example, Williams and Becklund~\cite{williams1989introduction}. Independent of the specific camera used, the MTF clearly states that coarser frequency levels contain more contrast. Consequently, any observation closer to the imaged object should be superior to other observations taken from farther distances. As our outlier removal accounts for unfocused images and misaligned image regions (see Section~\ref{sec:progressive-refinement:clean-observation}), we simply set the observation's confidence values $c^l_\mathcal{I}(x,y)$ to level map values of $\mathcal{L}$ and replace corresponding pixels on all Laplacian model levels, \ie
\begin{align*}
  \mathcal{M}^l(x,y) &\leftarrow 
  \begin{cases}
    \mathcal{I}^{l}(x,y) & \text{if } c^l_{\mathcal{I}}(x,y)  <
    c^l_{\mathcal{M}}(x,y),\\
    \mathcal{M}^l(x,y) & \text{else}.
  \end{cases}\\
  c^l_{\mathcal{M}}(x,y) &\leftarrow \min\{c^l_{\mathcal{I}}(x,y),\,
    c^l_{\mathcal{M}}(x,y)\}.
\end{align*}
This operation guarantees that the model stores the observation closest to the scene on a per-pixel level, \ie the model contains a single and reliable observation with maximal contrast. As we replace the model frequencies also on coarser Laplacian levels, we retain a photometrically and geometrically consistent reconstruction without any further post-processing.

{\bfseries Remark} Our choice of replacing frequencies instead of blending them is mainly motivated by the goal of being able to fuse several hundred images without global optimization.  We evaluated several blending strategies that have been able to retain fine geometric details for a small set of input images, but our experiments revealed that slight misalignments and improper masks lead to gradually increasing blur when applied to larger images sets.  Due to the non-perfect nature of image registration, blending all observations will wash out geometric details that will never be fully recovered by further blending operations. See the supplementary material for a comparison.

%-------------------------------------------------------------------------
\subsection{Refinement guidance}
\label{sec:progressive-refinement:visualize}  

\begin{figure}[t]
    \centering
    % the following command controls the width of the embedded PS file
    % (relative to the width of the current column)
    \includegraphics[width=1.0\linewidth]{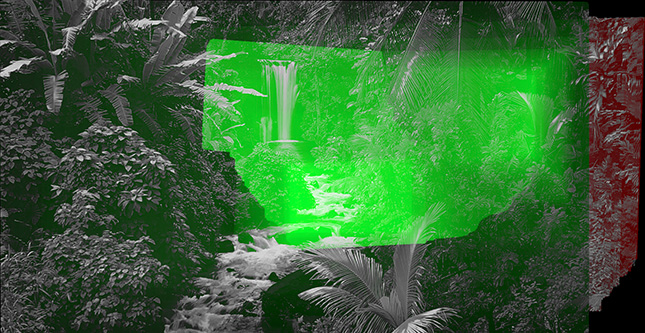}
    % replacing the above command with the one below will explicitly set
    % the bounding box of the PS figure to the rectangle (xl,yl),(xh,yh).
    % It will also prevent LaTeX from reading the PS file to determine
    % the bounding box (i.e., it will speed up the compilation process)
    % \includegraphics[width=.95\linewidth, bb=39 696 126 756]{sampleFig}
    %
    \caption{Rendering the confidence map shows the so far refined
      areas (green). The brighter the green colour, the finer the
      available geometric detail (\ie the lower $l$ for which
      $\mathcal{M}^{l}(x,y)$ exists). Red areas indicate regions with
      potential photometric inconsistency.}
    \label{fig:refinement-guidance}
\end{figure}

After the refinement, we render our confidence model map in order to
make the user aware of the current model composition in terms of
geometric detail. Figure~\ref{fig:refinement-guidance} shows such a
visualization for an example refinement. Using this visual guidance,
the user can steer the acquisition process according to his or her
needs and interests. We also visualize areas in which the initial
scene area defined by the reference image $\mathcal{I}_0$ has been extended by
further observations, as in this regions, our approach does not
guarantee photometric consistency (red areas in
Figure~\ref{fig:refinement-guidance}).
    
%-------------------------------------------------------------------------
\section{Results}
\label{sec:results}  

We evaluate the quality and the robustness of our progressive refinement imaging approach using 26 data sets, from which eight are presented in the paper; the remaining data sets can be found in the supplementary material. The data sets consist of photos as well as videos, captured with 29 different camera models (plus 19 unknown cameras). For each record, the reference image $\mathcal{I}_0$ is locally refined by inserting additional images of the same scene taken closer to the object or by zooming.

We compared our approach to 18 state of the art photo stitching methods using a sequence of panorama photos captured with different zoom levels and with moderate illumination changes (data set \textit{Panorama}) as well as the data sets \textit{De\"{e}sis mosaic} and \textit{House of Neptune and Amphitrite mosaic}. These comparisons are available in the supplementary material. Most of these methods fail to process the data sets properly and we observe the following behaviours:
\begin{enumerate*}[label=(\arabic*)]
\item The method reported that no matching of the input frames is possible.
\item The method did not achieve any refinement, \ie the merged image did not contain the fine details provided by the zoomed images. 
\item The method enforced a typical panorama scenario, resulting in a merged image, where the input images are aligned horizontally.
\end{enumerate*}

AutoStitch~\cite{brownAutoStitch,brown2007automatic} and Kolor Autopano Giga~\cite{kolorAutopano}, which is using the AutoStitch technology, were the only systems, able to reach a refinement. Unfortunately, AutoStitch crashes if the resolution of the merged image exceeds $30\;942$ pixels in one dimension. Furthermore, we had no access to Eisemann\etal's Photo~Zoom~\cite{eisemann2010photo}, which precludes experimental comparison.
   
In the following, we compare our approach to the unrefined input and the result of Autopano Giga~\cite{kolorAutopano}. To maintain the input images with the highest resolution in the final reconstruction, Autopano has to be operated using appropriate settings; see the supplementary material.

%-------------------------------------------------------------------------
\subsection{Refinement using different sources of imagery}
\label{sec:results:different-cameras}

For this experiment, we use photos that were captured from different sources on different dates using different cameras from various unknown positions. We use publicly available photos,~\eg from Flickr or Wikimedia Commons, which are unedited and  labelled for reuse with modification by the author.

\begin{description}
\item[House of Neptune and Amphitrite mosaic:] A photo of the mosaic at the \textit{House of Neptune and Amphitrite} in Herculaneum captured with a Pentax Optio S7 by Johnboy Davidson~\cite{davidson2007Herculaneum} is refined using six additional close-up photos %~\cite{neptune_2,neptune_3,neptune_4,neptune_5,neptune_6,neptune_7} 
captured with six different cameras (FUJIFILM FinePix F900EXR, Panasonic DMC-ZS6, Nikon D7100, 3 unknown cameras) in the years 2007, 2006, 2014, 2011, 2017, 2014 and 2009, respectively (see Figure~\ref{fig:results:houseOfNeptune}).
\end{description}

This data set comprises challenging illumination variations due to different camera hardware and post-processing. Feeding this data set into Autopano Giga results in a geometric consistent, but photometric inconsistent image, as Autopano Giga tries to generate smooth transitions between the individual photos. In contrast, our method yields photometric and geometric consistency.

\subsection{Robustness evaluation}
\label{sec:results:robustness-evaluation}

In this section, we compare our method to Autopano Giga under varying conditions regarding illumination (Section~\ref{sec:results:robustness-evaluation:illumination-changes}) and geometric consistency (Section~\ref{sec:results:robustness-evaluation:inconsistent-geometry}).

%-------------------------------------------------------------------------
\subsubsection{Inconsistent illumination}
\label{sec:results:robustness-evaluation:illumination-changes}

\begin{figure*}[h]
  \centering\scalebox{0.79}{
      \begin{tabular}{@{}l@{}}
          \begin{tabular}{ll@{}}
            \vspace{-2mm}
            \rotatebox{90}{\hspace*{2.5mm}9 photos} &
            \begin{subfigure}[t]{.793\linewidth}
              \includegraphics[width=1.0\textwidth]{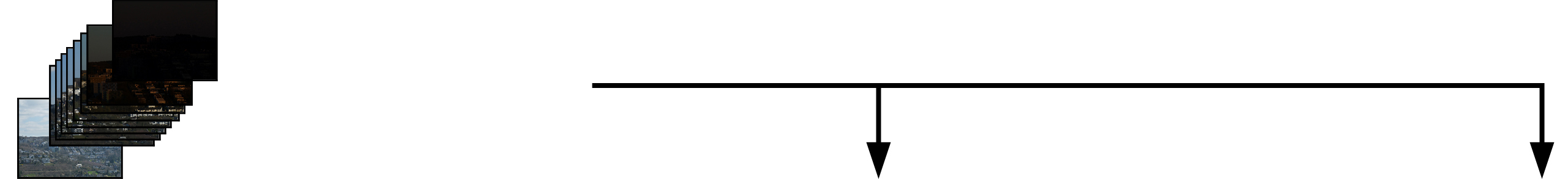}
              \label{fig:results:Panorama_at_different_daytimes.input_frames}
            \end{subfigure}
          \end{tabular} \\  
          \begin{subfigure}[t]{.33\linewidth}
            \includegraphics[width=0.99\textwidth, cfbox=red 1pt 0pt]{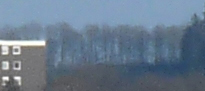}
          \end{subfigure}
          \begin{subfigure}[t]{.33\linewidth}
            \includegraphics[width=0.99\textwidth, cfbox=red 1pt 0pt]{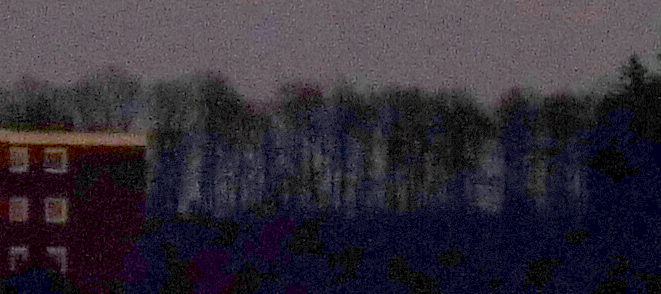}
          \end{subfigure}
          \begin{subfigure}[t]{.33\linewidth}
            \includegraphics[width=0.99\textwidth, cfbox=red 1pt 0pt]{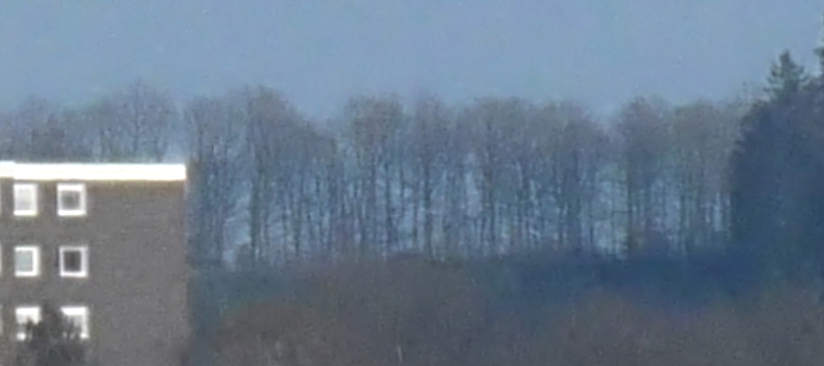}          
          \end{subfigure}\\
          \begin{subfigure}[t]{.33\linewidth}
            \includegraphics[width=1.0\textwidth]{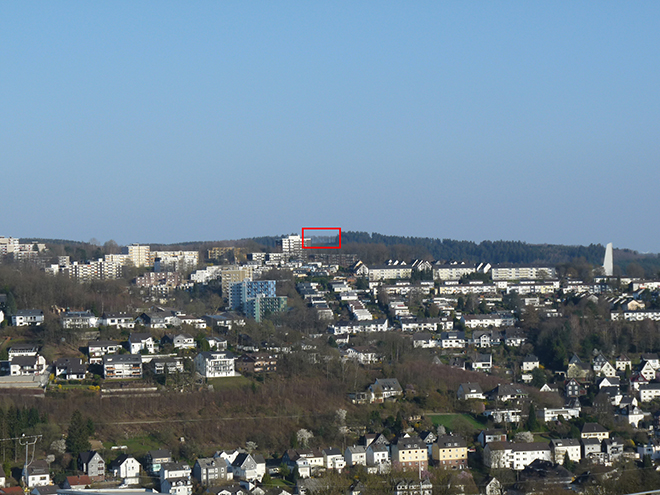}
            \caption{Reference image $\mathcal{I}_0$}
            \label{fig:results:Panorama_at_different_daytimes.input_1}
          \end{subfigure}
          \begin{subfigure}[t]{.33\linewidth}
            \includegraphics[width=1.0\textwidth]{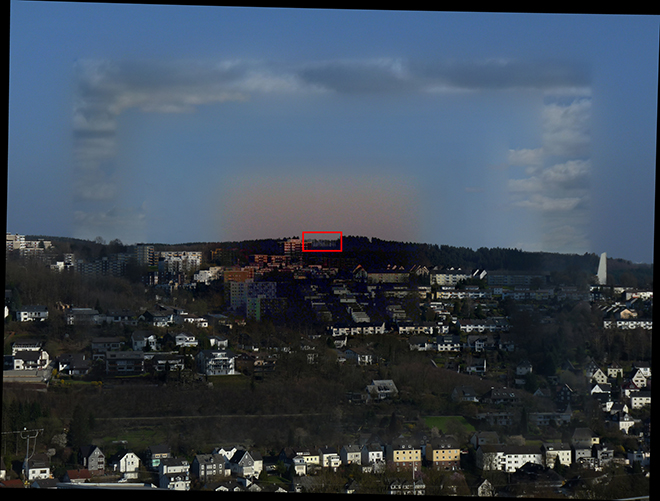}
            \caption{Autopano Giga}
            \label{fig:results:Panorama_at_different_daytimes.autopano}
          \end{subfigure}
          \begin{subfigure}[t]{.33\linewidth}
            \includegraphics[width=1.0\textwidth]{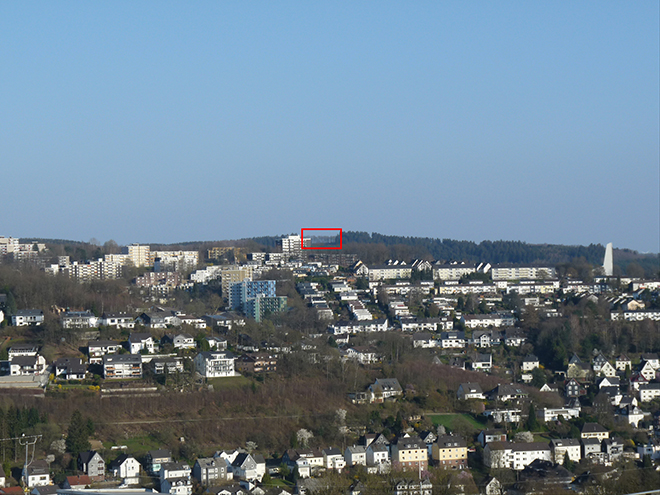}          
            \caption{Ours}
            \label{fig:results:Panorama_at_different_daytimes.ours}
          \end{subfigure}
      \end{tabular}
  }
  \caption{\textit{Panorama at different daytimes.}}
  \label{fig:results:Panorama_at_different_daytimes}
\end{figure*}

\begin{figure*}[h]
  \centering\scalebox{0.79}{
      \begin{tabular}{@{}l@{}}
          \begin{tabular}{ll@{}}
            \vspace{-2mm}
            \rotatebox{90}{\hspace*{3mm}38 photos} &
            \begin{subfigure}[t]{.793\linewidth}
              \includegraphics[width=1.0\textwidth]{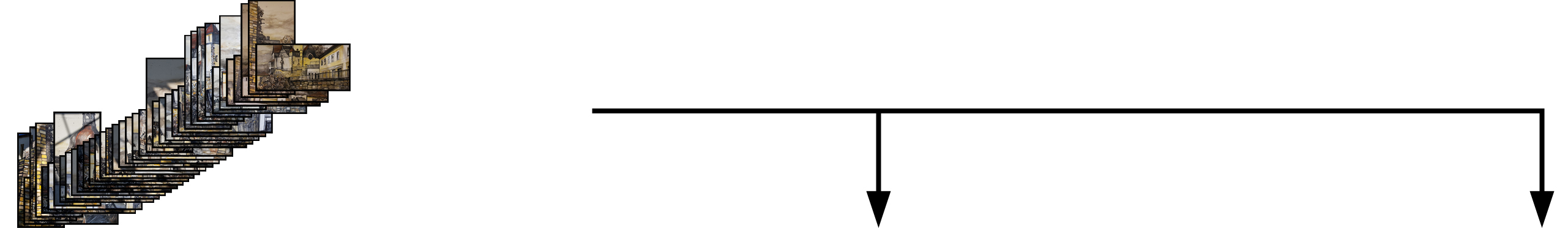}
              \label{fig:results:Wall_painting_Siegen_synagogue.input_frames}
            \end{subfigure}
          \end{tabular} \\  
          \begin{subfigure}[t]{.33\linewidth}
            \includegraphics[width=0.99\textwidth, cfbox=red 1pt 0pt]{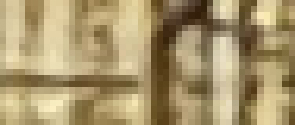}
          \end{subfigure}
          \begin{subfigure}[t]{.33\linewidth}
            \includegraphics[width=0.99\textwidth, cfbox=red 1pt 0pt]{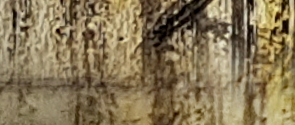}
          \end{subfigure}
          \begin{subfigure}[t]{.33\linewidth}
            \includegraphics[width=0.99\textwidth, cfbox=red 1pt 0pt]{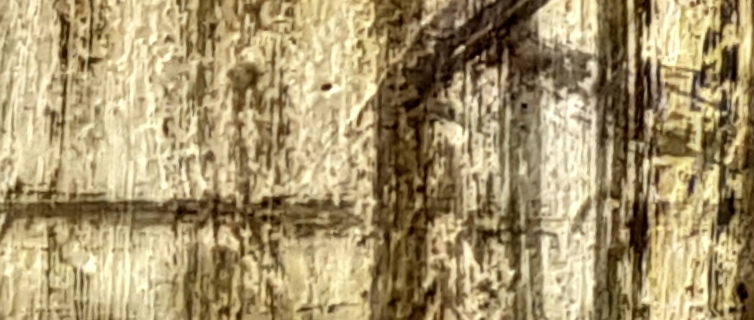}          
          \end{subfigure} \\
          \begin{subfigure}[t]{.33\linewidth}
            \includegraphics[width=1.0\textwidth]{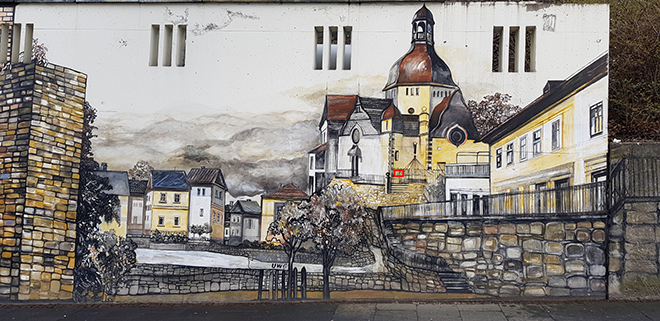}
            \caption{Reference image $\mathcal{I}_0$}
            \label{fig:results:Wall_painting_Siegen_synagogue.input_1}
          \end{subfigure}
          \begin{subfigure}[t]{.33\linewidth}
            \includegraphics[width=1.0\textwidth]{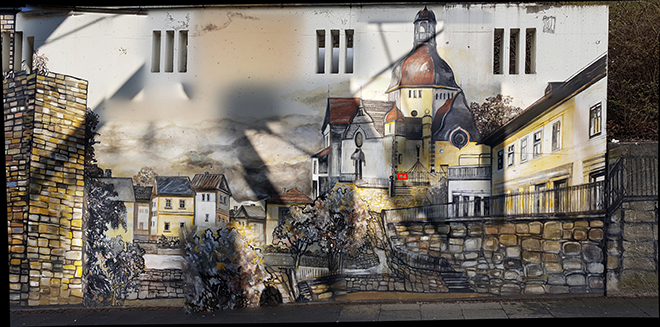}
            \caption{Autopano Giga}
            \label{fig:results:Wall_painting_Siegen_synagogue.autopano}
          \end{subfigure}
          \begin{subfigure}[t]{.33\linewidth}
            \includegraphics[width=1.0\textwidth]{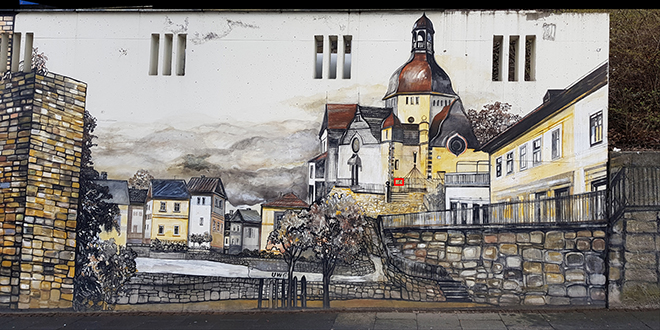}          
            \caption{Ours}
            \label{fig:results:Wall_painting_Siegen_synagogue.ours}
          \end{subfigure}
      \end{tabular}
  }
  \caption{\textit{Wall painting at different daytimes.}}
  \label{fig:results:Wall_painting_Siegen_synagogue}
\end{figure*}

\begin{figure*}[h]
  \centering\scalebox{0.79}{
      \begin{tabular}{@{}l@{}}
          \begin{tabular}{ll@{}}
            \rotatebox{90}{\hspace*{9mm}848 frames} &
            \begin{subfigure}[t]{.793\linewidth}
              \hspace{1.25mm}\includegraphics[width=0.985\textwidth]{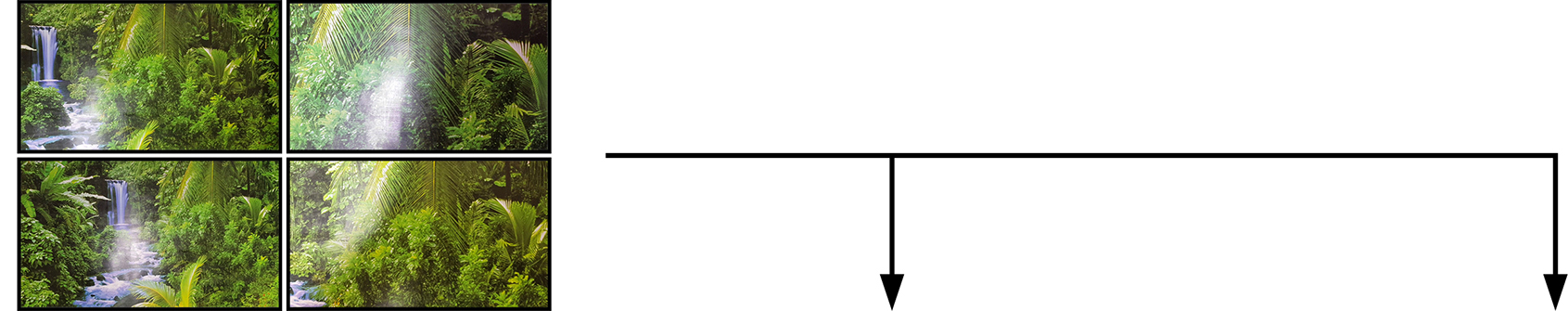}
              \label{fig:results:glossyPoster.input_frames}
            \end{subfigure}
          \end{tabular} \\  
          \vspace{-3mm}\begin{subfigure}[t]{.33\linewidth}
            \includegraphics[width=0.99\textwidth, cfbox=red 1pt 0pt]{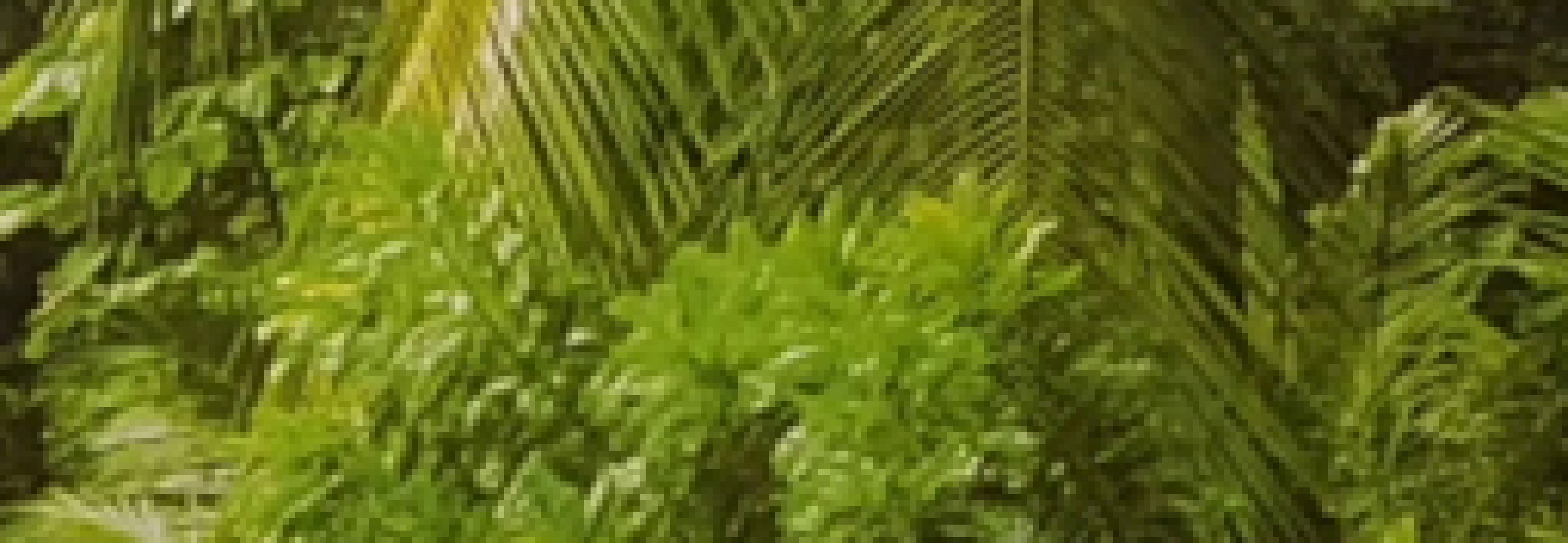}
            \label{fig:results:glossyPoster:closeup_reflection.input_1}
          \end{subfigure}
          \begin{subfigure}[t]{.33\linewidth}
            \includegraphics[width=0.99\textwidth, cfbox=red 1pt 0pt]{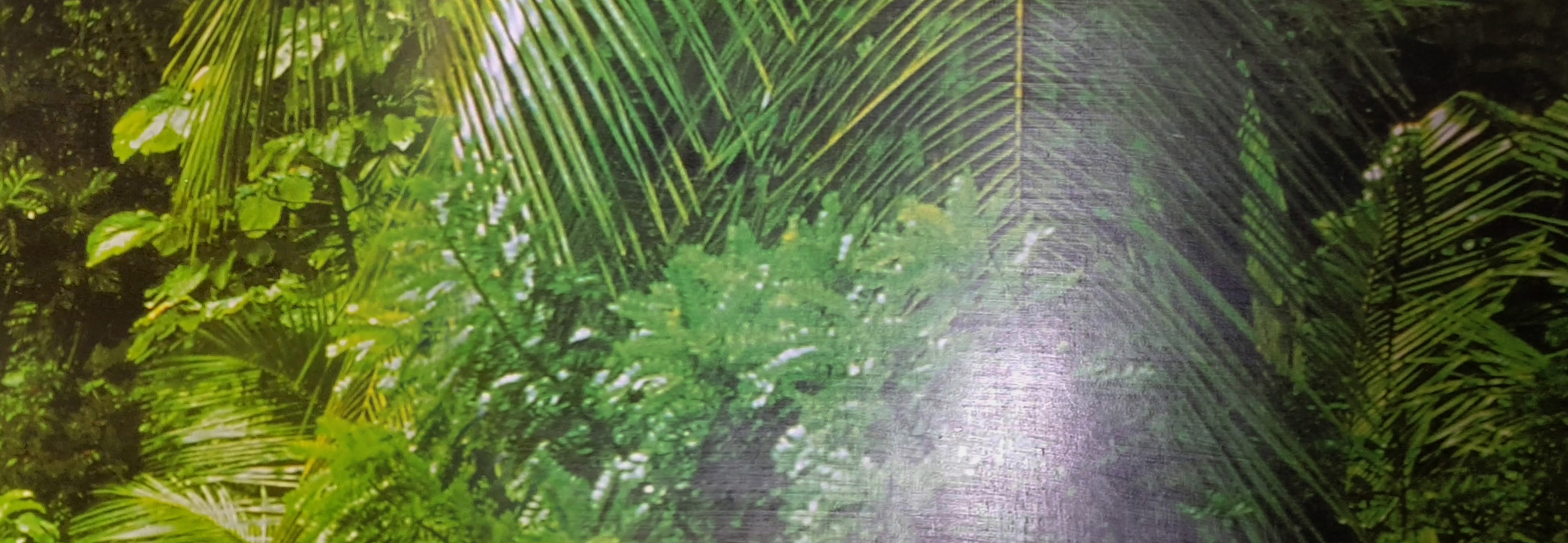}
            \label{fig:results:glossyPoster:closeup_reflection.autopano}
          \end{subfigure}
          \begin{subfigure}[t]{.33\linewidth}
            \includegraphics[width=0.99\textwidth, cfbox=red 1pt 0pt]{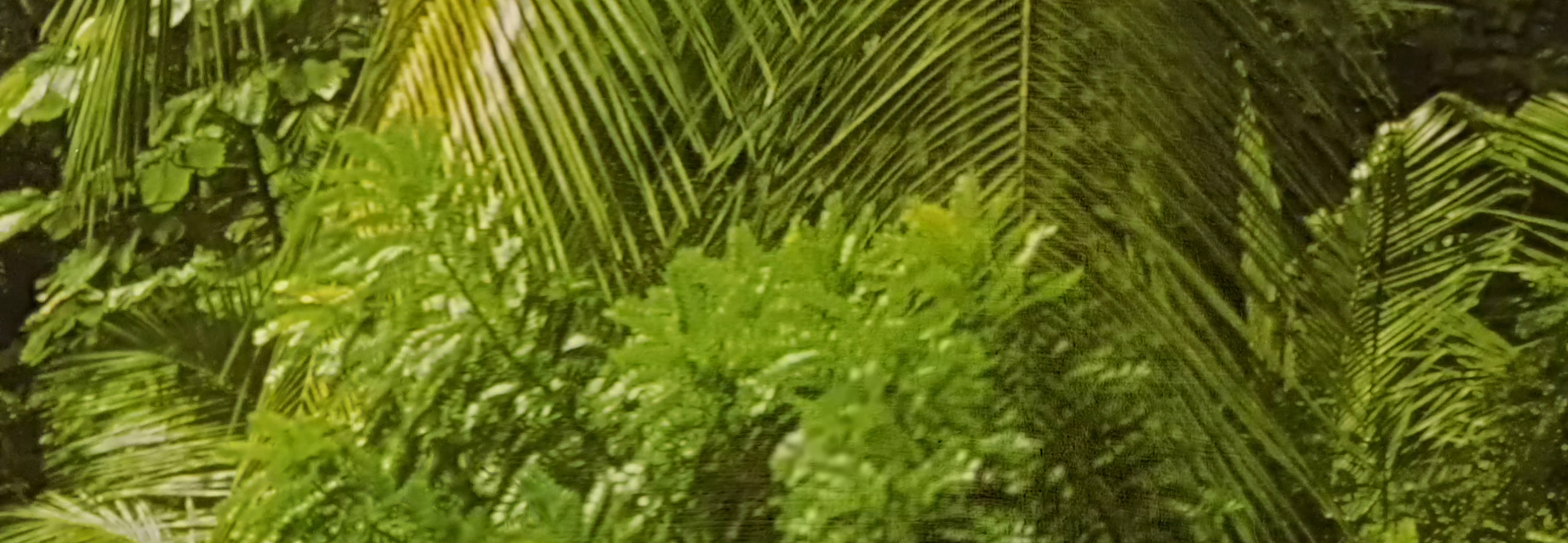}         
            \label{fig:results:glossyPoster:closeup_reflection.ours}
          \end{subfigure}\\
          \begin{subfigure}[t]{.33\linewidth}
            \includegraphics[width=1.0\textwidth]{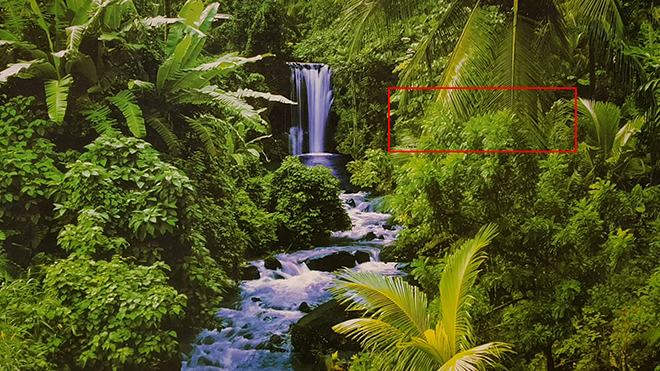}
            \caption{Reference image $\mathcal{I}_0$}
            \label{fig:results:glossyPoster.input_1}
          \end{subfigure}
          \begin{subfigure}[t]{.33\linewidth}
            \includegraphics[width=1.0\textwidth]{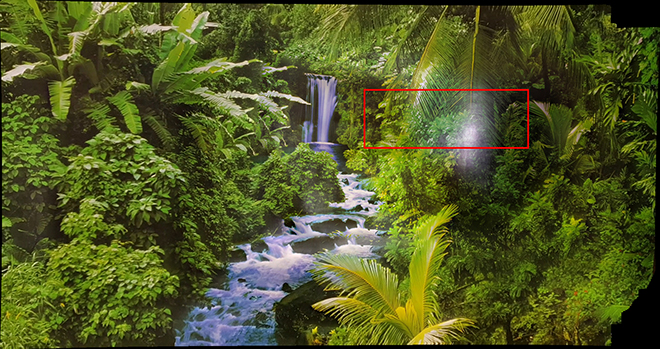}
            \caption{Autopano Giga}            
            \label{fig:results:glossyPoster.autopano}
          \end{subfigure}
          \begin{subfigure}[t]{.33\linewidth}
            \includegraphics[width=1.0\textwidth]{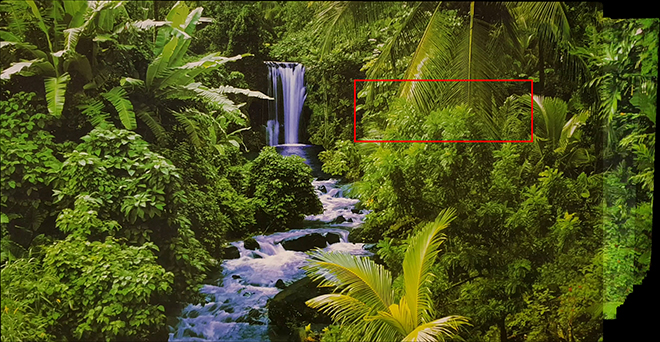} 
            \caption{Ours}                    
            \label{fig:results:glossyPoster.ours}
          \end{subfigure}
      \end{tabular}
  }
    \caption{\textit{Glossy poster}: The four sample frames (top) are part of the input video sequence, showing that the clip contains strong reflections.}
  \label{fig:results:glossyPoster}
\end{figure*}

\begin{figure*}[h]
  \centering\scalebox{0.79}{
      \begin{tabular}{@{}l@{}}
            \begin{tabular}{ll@{}}
            \vspace{-2mm}
            \rotatebox{90}{\hspace*{3mm}9 photos} &
            \begin{subfigure}[t]{.793\linewidth}
              \includegraphics[width=1.0\textwidth]{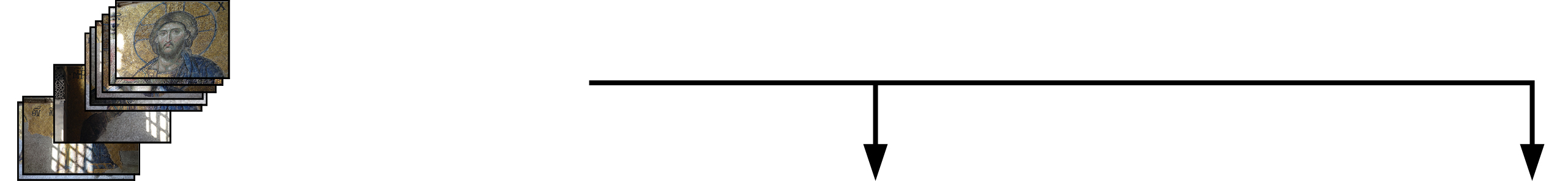}
              \label{fig:results:Deesis_mosaic.input_frames}
            \end{subfigure}
          \end{tabular} \\  
          \begin{subfigure}[t]{.33\linewidth}
            \includegraphics[width=0.99\textwidth, cfbox=red 1pt 0pt]{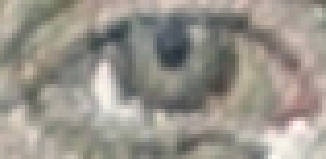}
          \end{subfigure}
          \begin{subfigure}[t]{.33\linewidth}
            \includegraphics[width=0.99\textwidth, cfbox=red 1pt 0pt]{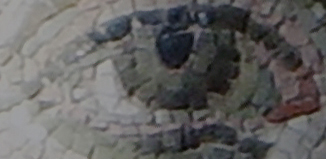}
          \end{subfigure}
          \begin{subfigure}[t]{.33\linewidth}
            \includegraphics[width=0.99\textwidth, cfbox=red 1pt 0pt]{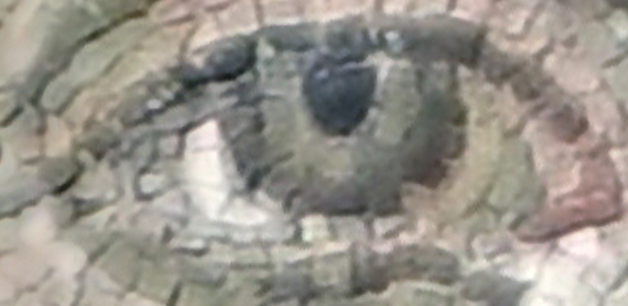}          
          \end{subfigure}\\
          \begin{subfigure}[t]{.33\linewidth}
            \includegraphics[width=1.0\textwidth]{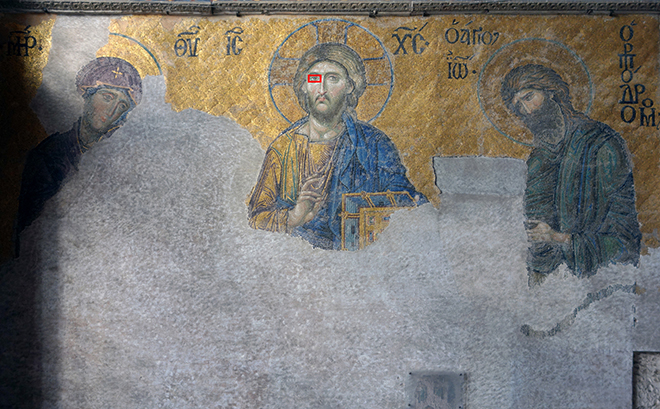}
            \caption{Reference image $\mathcal{I}_0$}
            \label{fig:results:Deesis_mosaic.input_1}
          \end{subfigure}
          \begin{subfigure}[t]{.33\linewidth}
            \includegraphics[width=1.0\textwidth]{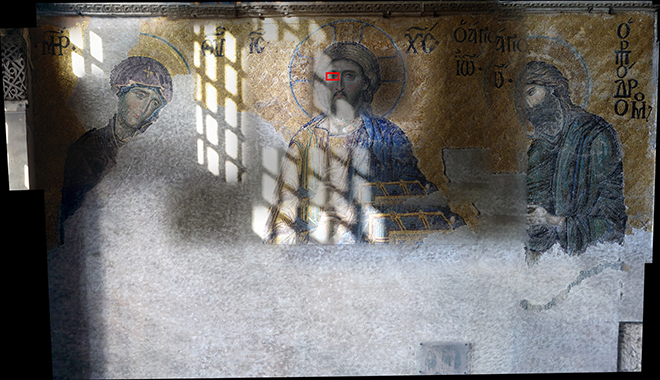}
            \caption{Autopano Giga}
            \label{fig:results:Deesis_mosaic.autopano}
          \end{subfigure}
          \begin{subfigure}[t]{.33\linewidth}
            \includegraphics[width=1.0\textwidth]{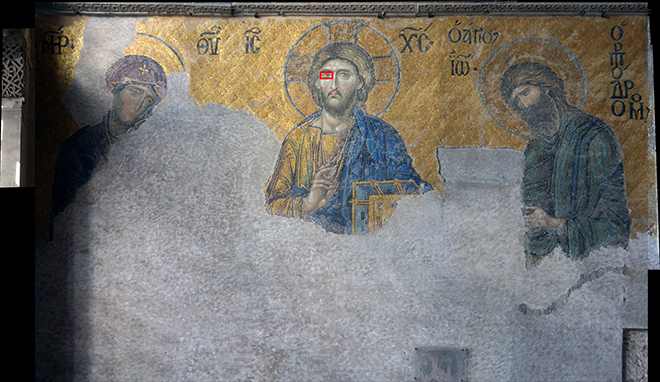}
            \caption{Ours}
            \label{fig:results:Deesis_mosaic.ours}
          \end{subfigure}
      \end{tabular}
  }
  \caption{\textit{De\"{e}sis mosaic.}}
  \label{fig:results:Deesis_mosaic}
\end{figure*}

The robustness against illumination changes is evaluated using the following four data sets:
\begin{description}
\item[Panorama at different daytimes:] A panorama shot is refined using nine additional zoomed-in photos that were taken at different daytimes with approximately 1~h time difference in the afternoon,
%at 01:00 pm, 02:00 pm, 03:00 pm, 04:00 pm, 05:00 pm, 06:00 pm, 07:00 pm, 07:30 pm, 08:00 pm
showing the same scene with decreasing sunlight, locally changing shadows and clouds, and with a fixed camera position (see Figure~\ref{fig:results:Panorama_at_different_daytimes}). All photos were captured with a Panasonic DMC-FZ28 ($3648\times2736$ pixels mode).
\item[Wall painting at different daytimes:] A photo of an outside wall painting is refined using 38 additional photos that were taken at different daytimes during a single day,
%10:30 am, 12:30 am, 02:30 pm, 05:30 pm, 07:30 pm, 09:30 pm, 
showing the same scene with varying sunlight and locally changing shadows on the wall from strongly varying camera poses (see Figure~\ref{fig:results:Wall_painting_Siegen_synagogue}). All photos were captured with a Samsung~Galaxy~S8 build-in camera ($4032\times 1960$ pixels mode).
\item[Glossy poster:] The first frame of a video sequence capturing a glossy poster is refined using the remaining 847 frames that were captured closer to the scene (every other frame of a 57~s video clip). The video was acquired with a Samsung~Galaxy~S8 build-in camera in 1080p mode. This sequence comprises frames with very strong photometric inconsistencies in terms of reflections (see Figure~\ref{fig:results:glossyPoster}).
\item[De\"{e}sis mosaic:] An overview photo of the Mosaic of the De\"{e}sis in the Hagia Sophia captured by Steven Zucker~\cite{zucker2012Deesis} is refined using nine additional close-up photos,
%~\cite{deesis_2,deesis_3,deesis_4,deesis_5,deesis_6,deesis_7, deesis_8,deesis_9,deesis_10}, 
where sunlight passes through the windows, resulting in a pattern of differently illuminated areas. All photos were captured with a Sony DSC-RX100 (see Figure~\ref{fig:results:Deesis_mosaic}).
\end{description}

%-------------------------------------------------------------------------
\paragraph*{Global illumination changes.} 

The first two data sets, \ie \textit{Panorama at different daytimes} (Figure~\ref{fig:results:Panorama_at_different_daytimes}) and \textit{Wall painting at different daytimes} (Figure~\ref{fig:results:Wall_painting_Siegen_synagogue}), contain major changes in global illumination, while \textit{Panorama at different daytimes} additionally contains geometric inconsistencies due to changes in cloudiness. While Autopano Giga has major difficulties in handling the illumination changes, the geometric variations (\textit{Panorama at different daytimes}) and the different camera poses (\textit{Wall painting at different daytimes}), our approach is able to combine both data sets into a photometric and geometric consistent image. The provided close-ups of the refined images demonstrate the proper handling of photometric and geometric information of our method during progressive image refinement.

%-------------------------------------------------------------------------
\paragraph*{Local illumination changes.} 

The second two data sets, \ie \textit{Glossy poster} (Figure~\ref{fig:results:glossyPoster}) and \textit{De\"{e}sis mosaic} (Figure~\ref{fig:results:Deesis_mosaic}), contain strong local illumination variations due to photoflash reflections and shadow casts by a window grating, respectively. In both scenarios, Autopano Giga is incorporating local illumination constellations from different close-up images into the reconstruction, resulting in very inconsistent intensity distributions in the output image. Our proposed progressive method is able to generate a photometric consistent result even under these extreme lighting conditions (see also Figure~\ref{fig:refinement-guidance} for a visualization of the refined areas for the \textit{Glossy poster} data set).

%Fig.~\ref{fig:results:glossyPoster} shows the \textbf{Glossy poster} scenario that comprises very strong illumination changes. 


%-------------------------------------------------------------------------
\subsubsection{Inconsistent scene geometry}
\label{sec:results:robustness-evaluation:inconsistent-geometry}

\begin{figure*}[h]
  \centering\scalebox{0.79}{
      \begin{tabular}{@{}l@{}}
          \begin{tabular}{ll@{}}
            \vspace{-2mm}
            \rotatebox{90}{\hspace*{3mm}2 photos} &
            \begin{subfigure}[t]{.933\linewidth}
              \includegraphics[width=1.0\textwidth]{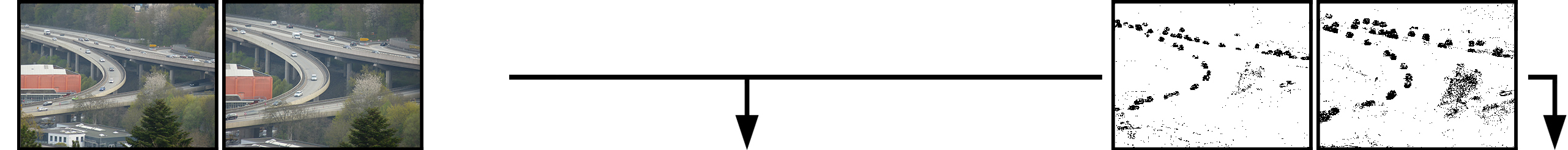}
              \label{fig:results:Moving_Cars.input_frames}
            \end{subfigure}
          \end{tabular} \\  
          \begin{subfigure}[t]{.33\linewidth}
            \includegraphics[width=0.99\textwidth, cfbox=red 1pt 0pt]{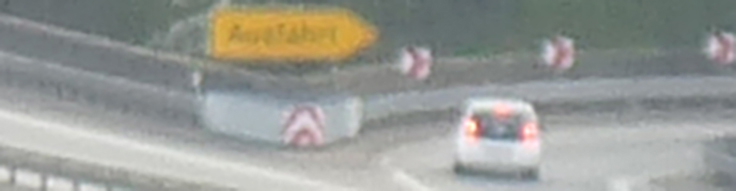}
          \end{subfigure}
          \begin{subfigure}[t]{.33\linewidth}
            \includegraphics[width=0.99\textwidth, cfbox=red 1pt 0pt]{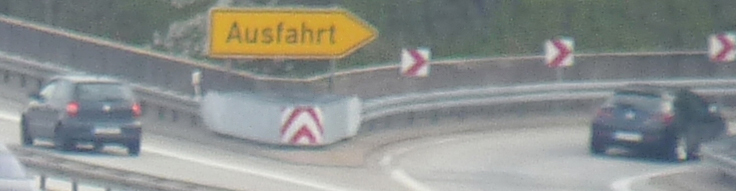}
          \end{subfigure}
          \begin{subfigure}[t]{.33\linewidth}
            \includegraphics[width=0.99\textwidth, cfbox=red 1pt 0pt]{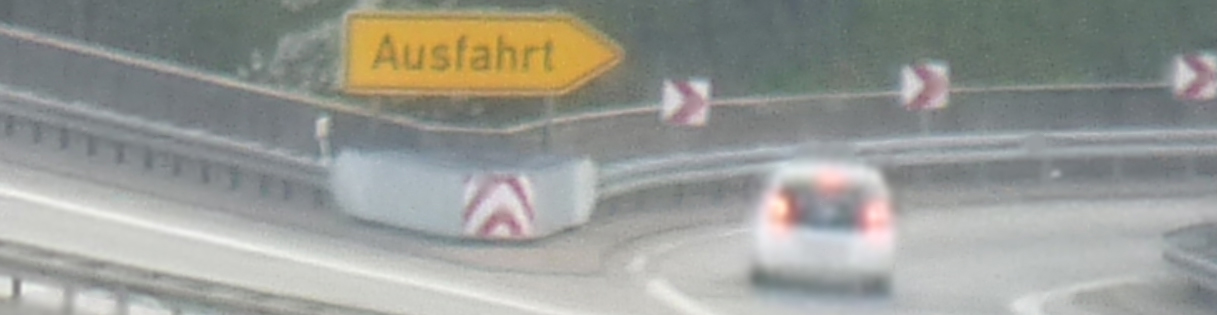}          
          \end{subfigure}\\
          \begin{subfigure}[t]{.33\linewidth}
            \includegraphics[width=1.0\textwidth]{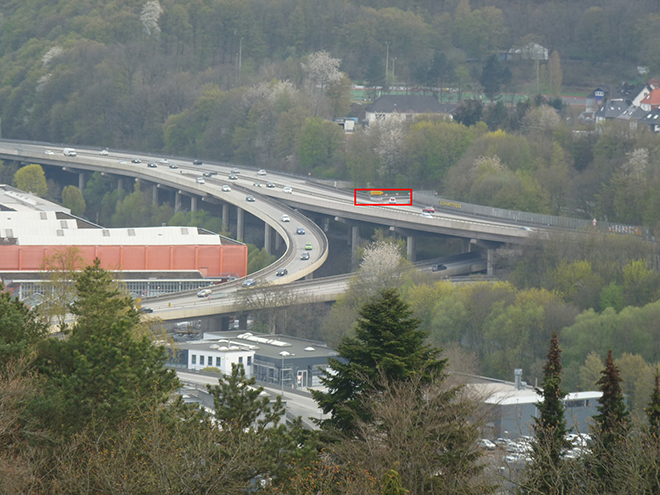}
            \caption{Reference image $\mathcal{I}_0$}
            \label{fig:results:Moving_Cars.input_1}
          \end{subfigure}
          \begin{subfigure}[t]{.33\linewidth}
            \includegraphics[width=1.0\textwidth]{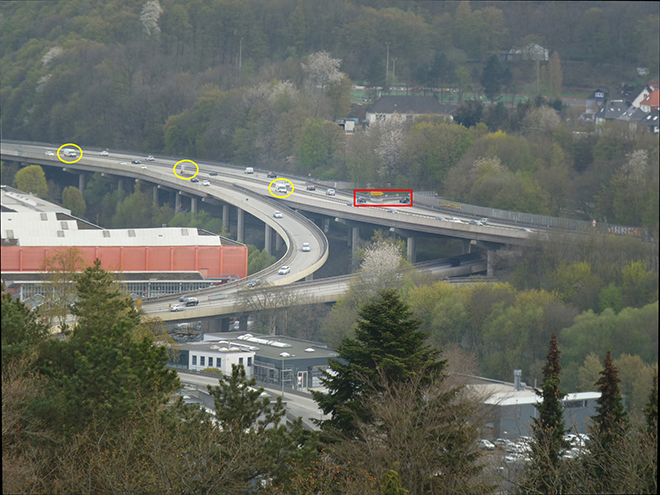}
            \caption{Autopano Giga}
            \label{fig:results:Moving_Cars.autopano}
          \end{subfigure}
          \begin{subfigure}[t]{.33\linewidth}
            \includegraphics[width=1.0\textwidth]{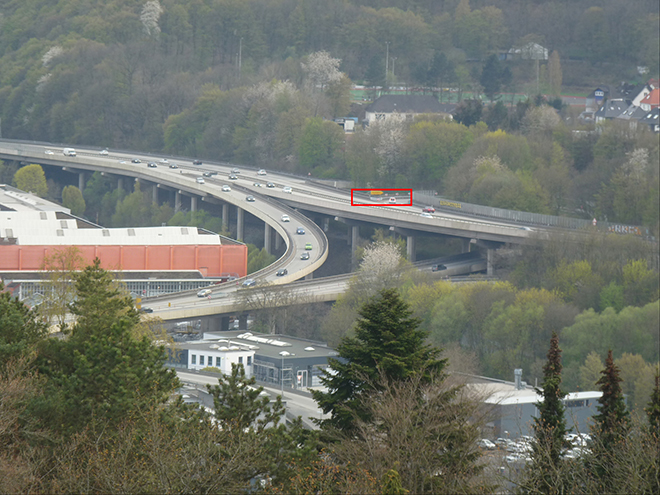}
            \caption{Ours}
            \label{fig:results:Moving_Cars.ours}
          \end{subfigure}
      \end{tabular}
  }
  \caption{\textit{Moving cars.}}
  \label{fig:results:Moving_Cars}
\end{figure*}

\begin{figure*}[h]
  \centering\scalebox{0.79}{
      \begin{tabular}{@{}l@{}}
          \begin{tabular}{ll@{}}
            \vspace{-2mm}
            \rotatebox{90}{\hspace*{1mm}1 photo} &
            \begin{subfigure}[t]{.793\linewidth}
              \includegraphics[width=1.0\textwidth]{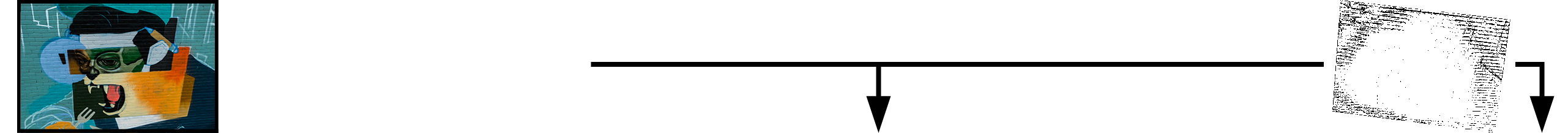}
              \label{fig:results:Streetart_Fisheye.input_frame}
            \end{subfigure}
          \end{tabular} \\
          \begin{subfigure}[t]{.33\linewidth}
            \includegraphics[width=0.99\textwidth, cfbox=red 1pt 0pt]{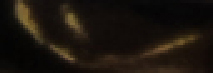}
          \end{subfigure}
          \begin{subfigure}[t]{.33\linewidth}
            \includegraphics[width=0.99\textwidth, cfbox=red 1pt 0pt]{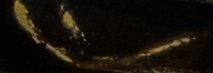}
          \end{subfigure}
          \begin{subfigure}[t]{.33\linewidth}
            \includegraphics[width=1.0\textwidth, cfbox=red 1pt 0pt]{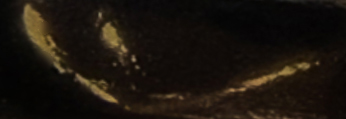}         
          \end{subfigure}\\
          \begin{subfigure}[t]{.33\linewidth}
            \includegraphics[width=1.0\textwidth]{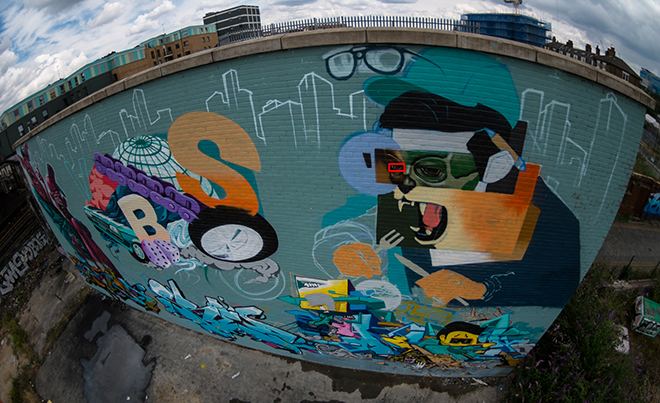}
            \caption{Reference image $\mathcal{I}_0$}
            \label{fig:results:Streetart_Fisheye.input_1}
          \end{subfigure}
          \begin{subfigure}[t]{.33\linewidth}
            \includegraphics[width=1.0\textwidth]{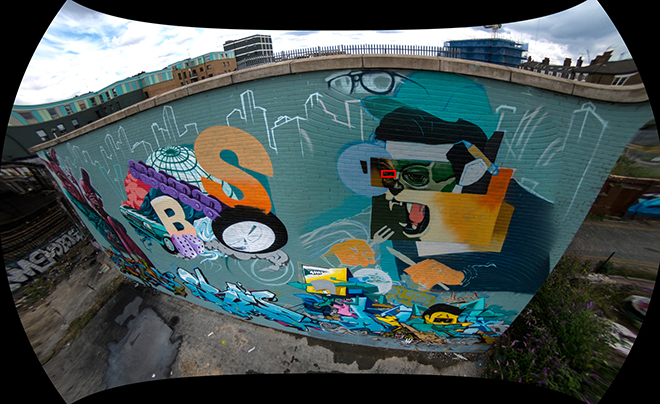}
            \caption{Autopano Giga}
            \label{fig:results:Streetart_Fisheye.autopano}
          \end{subfigure}
          \begin{subfigure}[t]{.33\linewidth}
            \includegraphics[width=1.0\textwidth]{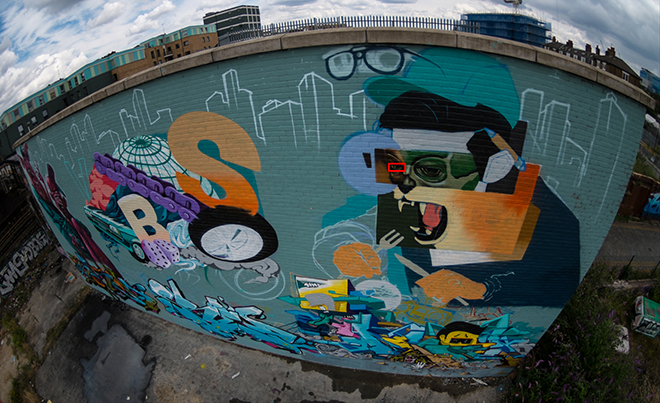}
            \caption{Ours}
            \label{fig:results:Streetart_Fisheye.ours}
          \end{subfigure}
      \end{tabular}
  }
  \caption{\textit{Streetart fisheye.}}
  \label{fig:results:Streetart_Fisheye}
\end{figure*}

The robustness against strong geometric variations is evaluated using the following two data sets:
\begin{description}
\item[Moving cars:] A panorama shot showing a freeway is refined using two additional zoomed-in photos, where the cars have been moving (see Figure~\ref{fig:results:Moving_Cars}). All photos were captured with a Panasonic DMC-FZ28 ($3648\times 2736$ pixels).
\item[Streetart fisheye:] An ultra-wide-angle shot of a street art graffito captured with an unknown camera with a fisheye lens by Mike Lambert~\cite{lambert2014Streetart} is refined using an additional photo captured with a normal lens (see Figure~\ref{fig:results:Streetart_Fisheye}).
\end{description}
We additionally depict the local outlier masks generated for both data sets; see Figures~\ref{fig:results:Moving_Cars} and \ref{fig:results:Streetart_Fisheye} and Section~\ref{sec:progressive-refinement:clean-observation}.

The main difference between both data sets is the type of geometric inconsistency. While the \textit{Moving cars} data set comprises locally unconstrained geometric variations, the \textit{Streetart fisheye} data set suffers from strong lens distributions that can be seen as globally constrained geometric inconsistencies. Both scenarios exhibit the different approaches taken by Autopano Giga and our method. While Autopano Giga generates visually pleasing output images in both cases, they both contain a mixture of all provided images leading to, \eg duplications of moving cars (see yellow circles in Figure~\ref{fig:results:Moving_Cars.autopano}) and a blended, deformed geometry in case of strongly varying lens artefacts (see Figures~\ref{fig:results:Streetart_Fisheye} and \ref{fig:results:outlier-removal}). In contrast, our method takes the initial image as photometric and geometric reference, and adjusts subsequent images to match this reference as closely as possible before adding details. Therefore, our approach delivers a consistent geometric result, \ie there are no multiple instances of moving objects or unexpected lens properties. Autopano Giga, however, always selects scene fragments with maximal focus, whereas our approach does not refine moving objects in the reference image, potentially leaving unsharp objects untouched; see Figure~\ref{fig:results:Moving_Cars.ours}. Consulting the local outlier masks, we can evaluate the overall quality of our two-stage registration process described in Section~\ref{sec:progressive-refinement:registration}; see also the discussion in Section~\ref{sec:results:pipeline-stages}. In the \textit{Moving cars} data set, mainly driving cars and moving trees are discarded and in the \textit{Streetart fisheye} data set, the strong lens distribution cannot be fully compensated by the optical flow stage.

\textbf{Remark} Image parallax due to non-planar scenes can be seen as a geometric inconsistency that is fixed by our local outlier removal. Consequently, image areas are not refined if the variation of the camera viewpoint leads to geometric inconsistencies due to strong depth inhomogeneities (see supplementary material). 

%-------------------------------------------------------------------------
%\paragraph*{Strong geometric discrepancy.}

%-------------------------------------------------------------------------
\subsection{Influence of pipeline stages}
\label{sec:results:pipeline-stages}

In the following, we discuss the influence of essential processing stages of our progressive image refinement pipeline; see Figure~\ref{fig:pipeline}. For this evaluation, we additionally use another data set:
\begin{description}
\item[Starlight:] A sequence of five photos captured free hand with a Samsung~Galaxy~S8 build-in camera with $1920\times1080$ pixels resolution, taken from an advertising poster.  
\end{description}

\begin{figure}[t!]
  \centering\scalebox{0.95}{
  \begin{subfigure}[t]{.49\linewidth}
    \includegraphics[width=1.0\textwidth]{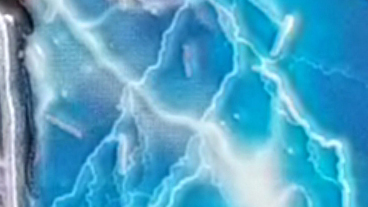}
    \caption{Without local fine-correction}
  \end{subfigure}
  \begin{subfigure}[t]{.49\linewidth}
    \includegraphics[width=1.0\textwidth]{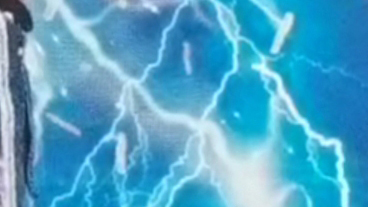}
    \caption{With local fine-correction}
  \end{subfigure}
    }
  \caption{A close-up comparison of the \textit{Starlight} data set without (left) and with locally refined image
    registration (right).}
  \label{fig:results:optical-flow}
\end{figure}

The \emph{Fine Registration} stage has a strong impact on the quality of the final result.  Figure~\ref{fig:results:optical-flow} demonstrates the effect of the locally refined image registration using optical flow on the \textit{Starlight} data set. Even for the comparable small lens distortion in this data set, we observe that the additional optical flow significantly improves the local matching of object details. This becomes even more apparent when images with strong optical distortions, such as the one in the \textit{Streetart fisheye}, are considered that cannot be modelled using a homography; see Figure~\ref{fig:results:Streetart_Fisheye}.

\begin{figure*}[t!]
  \centering\scalebox{0.85}{
      \begin{tabular}{@{}c@{}}
          \begin{subfigure}[t]{.33\linewidth}
            \includegraphics[width=1.0\textwidth]{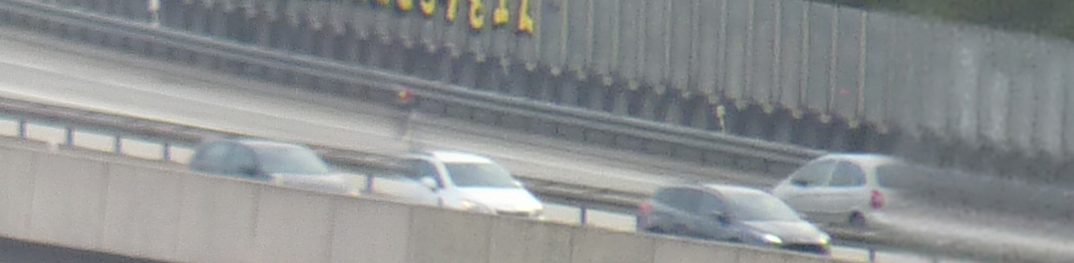}
          \end{subfigure}
          \begin{subfigure}[t]{.33\linewidth}
            \includegraphics[width=1.0\textwidth]{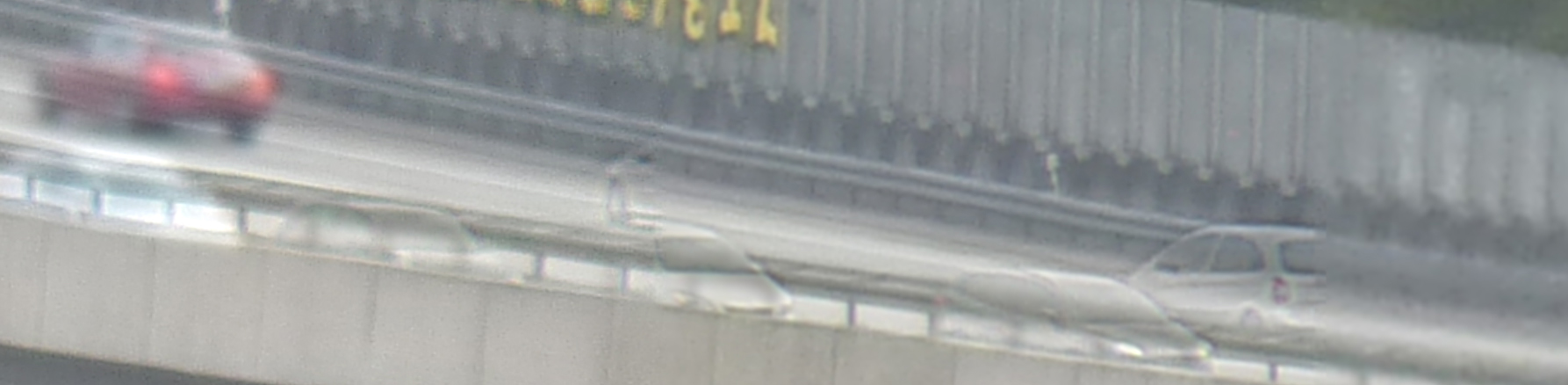}
          \end{subfigure}
          \begin{subfigure}[t]{.33\linewidth}
            \includegraphics[width=1.0\textwidth]{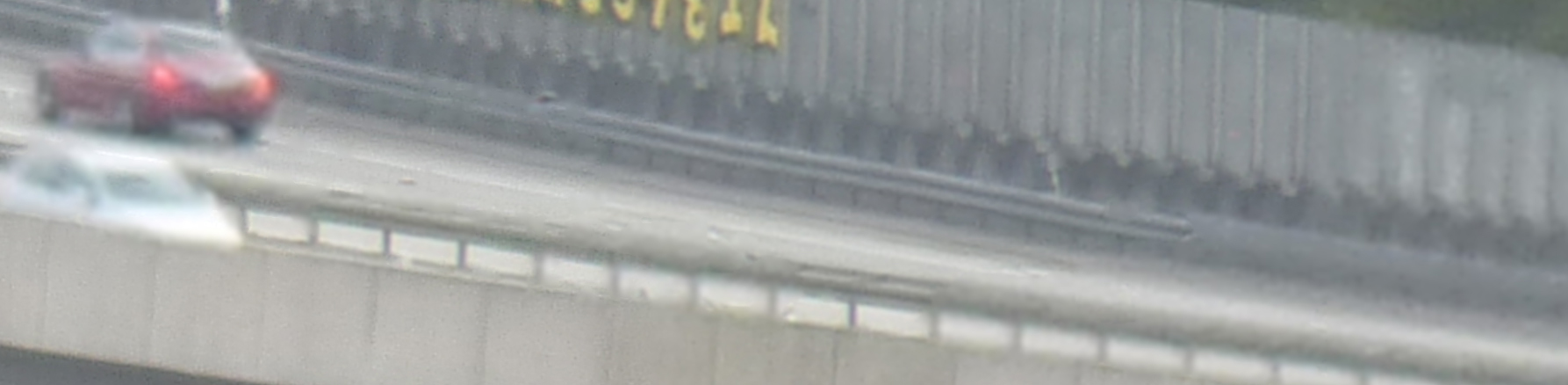}
          \end{subfigure} \\
          \begin{subfigure}[t]{.33\linewidth}
            \includegraphics[width=1.0\textwidth]{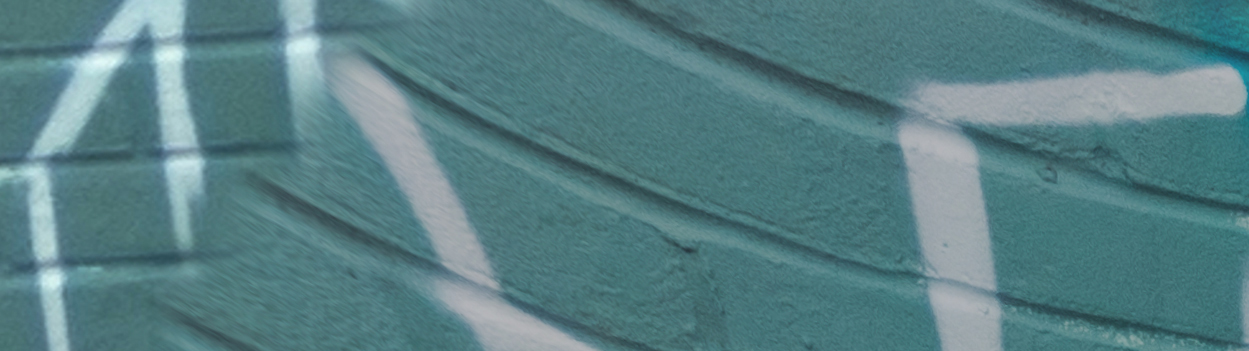}
            \caption{Autopano Giga}
          \end{subfigure}
          \begin{subfigure}[t]{.33\linewidth}
            \includegraphics[width=1.0\textwidth]{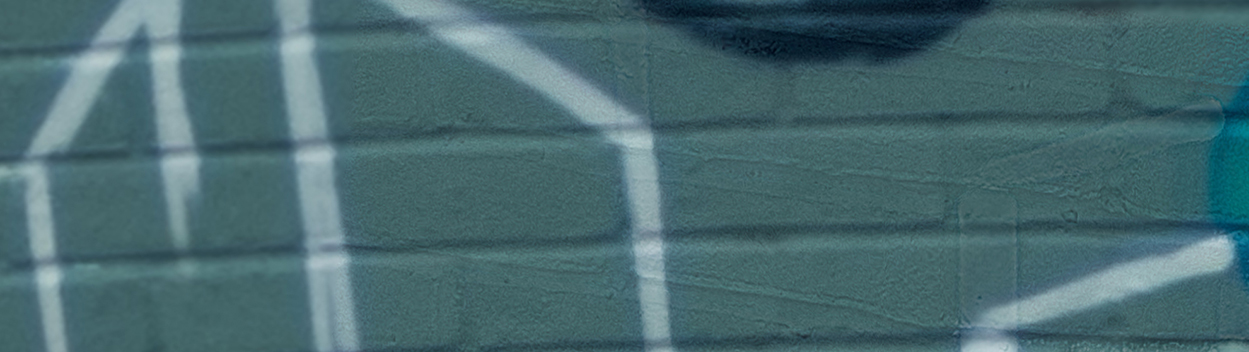}
            \caption{Ours without outlier removal}
          \end{subfigure}
          \begin{subfigure}[t]{.33\linewidth}
            \includegraphics[width=1.0\textwidth]{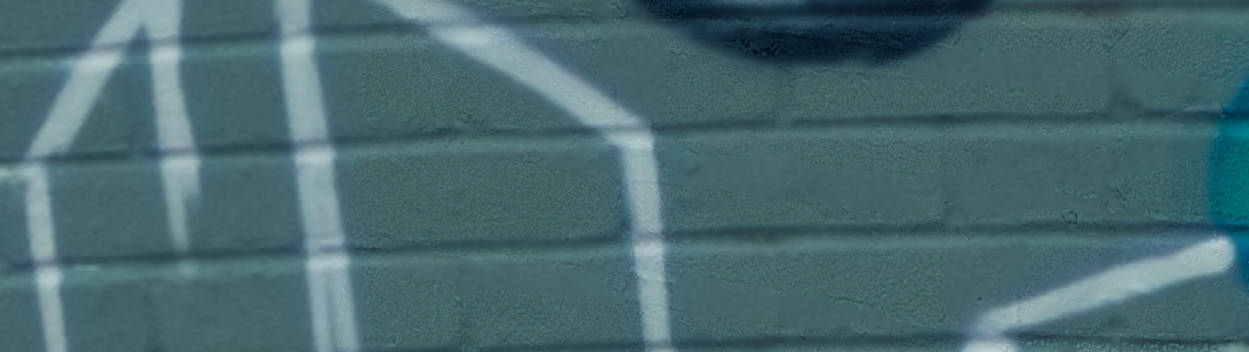}
            \caption{Ours with outlier removal}
          \end{subfigure}
      \end{tabular}
  }
  \caption{Influence of per-pixel outlier removal (top: \textit{Moving cars}, bottom: \textit{Streetart fisheye}).}
  \label{fig:results:outlier-removal}
\end{figure*}

\begin{table*}[t!]
    \small
    \centering
        \caption{Resources required for the complete refinement process, given for each data set with the number of input photos/pixels in total (AMD Ryzen Threadripper 1950X, 128 GB RAM, Nvidia Geforce GTX1080Ti). For the \textit{Glossy poster} data set, the timings per pipeline stage for our approach are: Image registration: 06:48 (min:s)/Pyramids generation: 03:23/Outlier removal: 01:20/Model expansion: $<00\text{:}01$/Merging Laplacian levels: 02:43.}
        \label{tab:panorama.resources}%
        \begin{tabular*}{\textwidth}{@{}l@{\extracolsep{\fill}}@{}S[table-format=3.2]@{}S[table-format=1.2]@{}r@{}c@{}c@{}r@{}}
        \toprule
        & \phantom{abc} & \multicolumn{2}{@{}c@{}}{Peak total RAM usage (GB)} & \phantom{abc} & \multicolumn{2}{@{}c@{}}{\vspace{0.5ex}Processing time (min:s)} \\ \cline{3-4}\cline{6-7} 
        & & {Autopano Giga} & Ours
        & & {Autopano Giga} & Ours \rule{0pt}{3.0ex}\\
        \midrule
        De\"{e}sis mosaic (10 photos/0.12 gigapixel) & & 31.16 & 5.12 & & {01}:{52} & {00}:{40} \\
        Glossy poster (848 frames/1.76 gigapixel) & & 121.53 & 2.23 & & {70}:{34} & {14}:{14} \\
        House of Neptune and Amphitrite mosaic (7 photos/0.01 gigapixel) & & 17.52 & 1.45 & & {01}:{25} & {00}:{06} \\
        Moving cars (3 photos/0.03 gigapixel) & & 4.15 & 2.00 & & {00}:{26} & {00}:{05} \\
        Panorama at different daytimes (10 photos/0.10 gigapixel) & & 14.97 & 2.57 & & {01}:{18} & {00}:{27} \\
        Streetart fisheye (2 photos/0.03 gigapixel) & & 7.38 & 2.52 & & {00}:{33} & {00}:{05} \\
        Wall painting at different daytimes (39 photos/0.31 gigapixel) & & 68.73 & 7.56 & & {06}:{13} & {02}:{35} \\
        \bottomrule
        \end{tabular*}%
\end{table*}

The effect of the \emph{Per-Frame Outlier Removal} is demonstrated in the \textit{Panorama at different daytimes} data set; see Figure~\ref{fig:results:Panorama_at_different_daytimes}. Here, the last input frame, which has been captured in very weak sunlight, has not passed the check, \ie it has been discarded for model image refinement, since it does not provide additional image details. In comparison, Autopano Giga performs a histogram equalization and incorporates the last frame, overwriting the details of the previous frames, which results in a loss of detail and increased noise in the refined image. For the \textit{Glossy poster} data set, 2.01\% of the input frames were rated unable to contribute finer details (full image outlier reject), hence only newly observed areas were incorporated into the model if available.
The \emph{Per-pixel Outlier Removal} as described in Section~\ref{sec:progressive-refinement:clean-observation} is evaluated in Figure~\ref{fig:results:outlier-removal}, which contains close-ups of the \textit{Moving cars} and \textit{Streetart fisheye} scenarios, for which we lowered the threshold for discarding pixels to $E(x,y)>1$. Deactivating the local outlier removal yields artefacts visible as slight ghosting of cars and of mismatching seams in the \textit{Moving cars} and \textit{Streetart fisheye} scenarios, respectively. Both effects vanish nearly completely if the per-pixel outlier removal gets activated. 

%-------------------------------------------------------------------------
\subsection{Comparison of required resources}
\label{sec:results:required-resources}

Table~\ref{tab:panorama.resources} shows for each data set a comparison of peak total RAM usage and processing time for the whole refinement process for both Autopano Giga and our proposed method. This comparison demonstrates that global optimization significantly increases memory requirements and runtime. This is unavoidable as global optimization methods have to keep all relevant images in memory in order to process them jointly. Especially for the video data set \textit{Glossy poster}, the memory requirements increase severely by a factor of approximately 40, whereas the processing time increases by a factor of 5. In contrast, our approach of progressively refining the image is much more lightweight and continuously eliminates redundancy, substantially lowering resource requirements.

In our implementation, we mainly optimized our adaptive Laplacian pyramid as described in Section~\ref{sec:progressive-refinement}, while the main image processing stages, such as feature extraction, optical flow and basic image operations, are taken from OpenCV as is.

%-------------------------------------------------------------------------
\subsection{Limitations and discussion}
\label{sec:results:discussion}  

Our current pipeline can guarantee photometric consistency only within the region of the scene observed by the initially captured reference frame $\mathcal{I}_0$. Our system is capable of incorporating images that are partially outside this initial region, but at the seam to $\mathcal{I}_0$, it yields geometric but no photometric consistency. Furthermore, since the refined image is always consistent to the reference image, unintended photometric effects in $\mathcal{I}_0$, \eg  photoflash reflections, will not be compensated by additional photos.
Moreover, our current implementation is not re-entrant, \ie it does not support the continuation of a previously acquired model image represented in a Laplacian pyramid as described in Section~\ref{sec:representation}. Although the implementation of this functionality is of some practical importance, we consider it an engineering task.
While the system is truly progressive, in that information is fed frame-by-frame without any global optimization (across several images), the current implementation is interactive but not real time. So far, we have not fully optimized and tightly integrated the pipeline components in order to achieve optimal load and compute balancing, \eg by leveraging concurrency. Apparently, faster executions of dense image processing operations, \eg optical flow, will have direct impact on the performance (see Table~\ref{tab:panorama.resources}).
Furthermore, the fine image registration using optical flow cannot correct strong optical distortions or parallax; however, our per-pixel outlier removal compensates for this error almost entirely; see Figure~\ref{fig:results:outlier-removal}.

%-------------------------------------------------------------------------
\section{Conclusions}
\label{sec:conclusions}  

We presented a simple, yet very effective and efficient technique for the progressive incorporation of large image sequences into a single, geometrically and photometrically consistent model image. Conceptually, our approach has no restriction to object resolution, camera-to-object distance, camera intrinsics or acquisition setup. Additionally, our approach does not require a global optimization applied to the complete input image set, or to parts thereof. Our approach achieves geometric registration using a two-stage approach that combines a homography and an additional local refinement using a flow field. It can handle strong illumination changes, yielding photometrically consistent results. Due to its progressive nature, our approach allows for a valid and consistent reconstruction at any moment during the refinement process without any post-processing.

%-------------------------------------------------------------------------

%\bibliographystyle{eg-alpha}
\bibliographystyle{eg-alpha-doi}
\balance
\bibliography{paper}

%-------------------------------------------------------------------------
\section*{Supporting Information}
Additional supporting information may be found online in the supplementary material.

\vspace{2mm}
\noindent \textbf{Figure 1:} A Burial at Ornans.

\noindent \textbf{Figure 2:} Amalfi cathedral.

\noindent \textbf{Figure 3:} Brandenburg Gate.

\noindent \textbf{Figure 4:} Coronation of Napoleon.

\noindent \textbf{Figure 5}: Dendera crypt relief.

\noindent \textbf{Figure 6:} Glorification of Saint Ignatius.

\noindent \textbf{Figure 7:} House at Lake Garda.

\noindent \textbf{Figure 8:} Lake Garda.

\noindent \textbf{Figure 9:} Pewter figures.

\noindent \textbf{Figure 10:} Raft of the Medusa.

\noindent \textbf{Figure 11:} Ship painting.

\noindent \textbf{Figure 12:} Streetart.

\noindent \textbf{Figure 13:} The Wedding Feast at Cana.

\noindent \textbf{Figure 14:} Villa of the Mysteries (back wall).

\noindent \textbf{Figure 15:} Villa of the Mysteries (left wall).

\noindent \textbf{Figure 16:} Winter scene in Brooklyn.

\noindent \textbf{Figure 17:} Starlight (five images): five photos captured with a Samsung Galaxy S8 build-in camera were merged.

\noindent \textbf{Figure 18:} Starlight (477 frames): comparison between the blending (middle) and the replacement merge strategy (right) applied to a 477 frames sequence captured with a Samsung Galaxy S8 build-in camera, downsampled to $960\times540$ pixels.

\noindent \textbf{Table 1}: Panorama: comparison to state-of-the-art photo stitching methods.

\noindent \textbf{Table 2:} De\"{e}sis mosaic: comparison to state-of-the-art photo stitching methods.

\noindent \textbf{Table 3:} House of Neptune and Amphitrite mosaic: comparison to state-of-the-art photo stitching methods.

%-------------------------------------------------------------------------

\end{document}